\documentclass[12pt,a4paper,twoside,final,notitlepage]{article}
\usepackage[english]{babel}
\usepackage[T1]{fontenc}  
\usepackage{graphicx}
\usepackage[fleqn, leqno]{amsmath}

\newcommand{\monitem}{ \smallskip \noindent $\bullet$ \quad  }
\newcommand{\smb} { {\scriptstyle  \bullet  } }
\newcommand{\moneq} {\vspace*{-18pt} \begin{equation} \displaystyle \qquad } 
\newcommand{\eqnono}{\vspace*{-25pt} \begin{equation*} \displaystyle } 
\newcommand{\nototo}{ \vspace*{-10pt} \noindent }
\def\sqr#1#2{{\vcenter{\vbox{\hrule height.#2pt \hbox{\vrule width .#2pt height#1pt 
\kern#1pt \vrule width.#2pt} \hrule height.#2pt}}}}
\def\square{\mathchoice\sqr64\sqr64\sqr{4.2}3\sqr33}
\def\tvi {\vrule  height 10pt depth 5pt width 0pt}
\def\tv  {\tvi \vrule}
\def\tvg {\tv ~~}
\def\tvd {~~ \tv}
\def \na{ \noalign {\hrule}  }
\def\hcr {\hfill & \cr}

\begin{document}
\bibliographystyle{alpha}

\title{ \vskip -2 cm  {  \bf \LARGE On lattice Boltzmann scheme, finite volumes \\
and boundary conditions }\\~ } 

\author { {\large   Fran\c{c}ois Dubois $^1$ $^2$ \small 
and \large  Pierre Lallemand $^3$ } \\ ~\\
{\it \small  $^1$  Conservatoire National des Arts et M\'etiers, Paris, France.}  \\
{\it  \small $^2$ Numerical Analysis and Partial Differential Equations }   \\
 {\it  \small  Department of Mathematics, Paris Sud  University,  Orsay, France. } \\
{\it  \small $^3$ Applications Scientifiques du Calcul Intensif, Orsay,  France.  } \\
{ \rm  \small francois.dubois@math.u-psud.fr, pierre.lal@free.fr}   }

\date { { \large  02 april  2008 } 
\footnote {\rm  \small $\,\,$ Presented at the Third
International Conference for Mesoscopic Methods in Engineering and Science, 
 Hampton,  25-28 July 2006 and  published in {\it Progress in Computational Fluid Dynamics}, 
volume 8, pages 11-24, 2008. Edition June 2023. }}

\maketitle
\renewcommand{\baselinestretch}{1.}
\bigskip 

\noindent {\it \bf \large Abstract } 

\noindent
We develop the  idea  that a natural link between Boltzmann
schemes and finite volumes exists naturally: the conserved mass and momentun
 during the collision phase of the Boltzmann scheme 
induces general  expressions for mass and momentum fluxes. 
We treat a unidimensional case and  
focus our development in two dimensions on possible flux boundary conditions. 
Several test cases show that a high level of accuracy can be achieved with this scheme.


\large

\section{Introduction}\label{intro}

\nototo \monitem
The lattice Boltzmann scheme is a popular numerical method based on a kinetic approach for
fluid dynamics  (\cite{hpp}  \cite{dlf} \cite{fhp}  \cite{mz}  
\cite{hsb} \cite{dh92}  \cite{kr95}  \cite{ll00}). An exact  propagation step 
in a lattice is followed
by a local relaxation process. It has been very early recognized (see {\it e.g.}
\cite{bsv05}) that the lattice Boltzmann scheme is compatible with mass and momentum
conservation. Similarly, classical conservation laws that lead  to finite volume
methods (see    {\it e.g.} \cite{pa80}, \cite{gr96} or \cite{dd05}) incorporate explicitly the
evaluation of numerical fluxes associated with conserved physical quantities. 
In order to extend the lattice Boltzmann scheme to unstructured meshes, several authors  
\cite{ch98}  \cite{pxdc99} \cite{ubs03} \cite{usb04} start from the kinetic equations for
the particle distribution and use control volumes ``\`a la INRIA''  \cite{adlv83}
 \cite{vi86}, that is control volumes around the vertices of the triangulation.

\smallskip  \monitem
On the other hand, the treatment of boundary conditions with the help of
 boundary fluxes is natural with the so-called cell centered version of the finite volume
 method (see {\it e.g.} the classical monograph of Roache  \cite{ro72} and our
 contributions   \cite{df89}   \cite{du01} in the strong nonlinear case). The incorporation 
of mass conservation {\it via} a zero mass flux on a solid boundary of the domain has been
studied by D'Humi\`eres   \cite{dh01} and also developed in   \cite{vds06} and   \cite{hhc}.

\smallskip  \monitem
 In what
follows, we start from a very general lattice Boltzmann scheme and propose to incorporate 
the fundamental conservations of mass and momentum in the framework of finite
 volumes. Then we propose to develop boundary conditions based on mass flux for the
 one-dimensional lattice Boltzmann scheme with three velocities. 
We extend the previous ideas  for the so-called D2Q9 two-dimensional model. We extend also
 these ideas to the treatment of boundary conditions where the geometry of the control
 volumes has to be modified in order to take into account the physical geometry. 
Numerical simulations show the interest of our approach.

\section{About the property of conservation}\label{volfi}

\nototo \monitem
We denote by   $\, \cal{ L} \, $   a  lattice, $\, \Delta x \,$ a typical scale associated
with this lattice, $\, \Delta t \,$ a time step, 

\moneq
\lambda \equiv  {{ \Delta x}\over{ \Delta t}} 
\label {lambda}  \end{equation}

\smallskip \noindent
a typical speed of the problem, $x$ a vertex of this lattice,

\moneq
x_j \equiv    x \,+\,  \Delta t  \, \,  v_j \,, \quad   0 \leq j \leq J \,,\, 
\label {sommet}  \end{equation}

\smallskip \noindent
the set of neighbouring nodes around the vertex $x$. 
Note that the node defined by the relation   (\ref{sommet}) 
is a vertex of the lattice.  We suppose that the family 
$\, ( v_j)_{ 0 \leq j \leq J}\,$   of speeds  is symmetric relative to the origin, 
as an example  is shown in Figure 1.  

\moneq
\forall j \, \in \{ 0,\,  \cdots,  J \}, \quad  \exists \, {\rm ! } \, 
\sigma(j)  \, \in \{ 0,\,  \cdots,  J \}, \quad  v_j + v_{\sigma(j)} \, = \, 0 \, . \,
\label{symetrie}  \end{equation}

\smallskip \noindent
We remark the clear property of involution:

\moneq
\sigma \big( \sigma(j) \big) = j \,, \qquad 0 \leq j \leq J \, . 
\label{sigma-2}  \end{equation}

\bigskip  
\begin{center}   \includegraphics[width=5cm,height=5cm]  {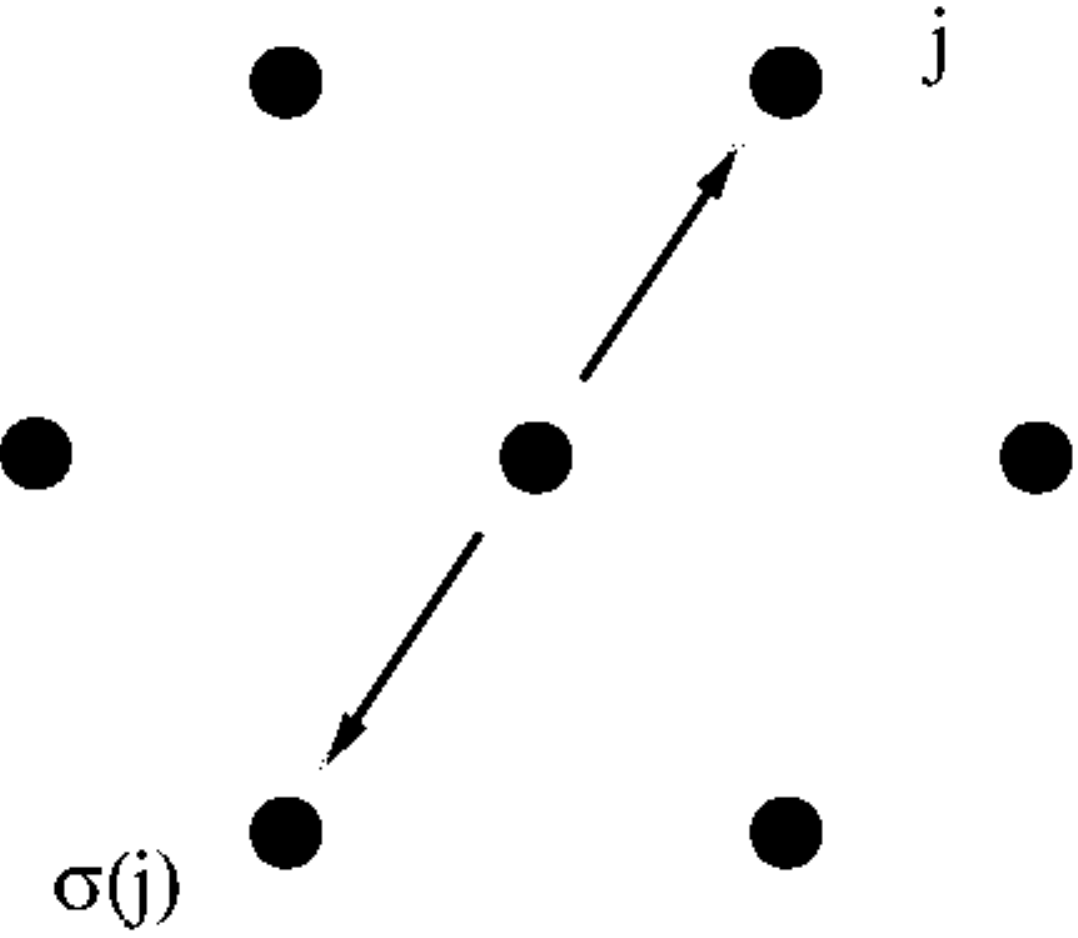} \end{center}
\begin{center} \it Figure 1. Central symmetry hypothesis  \end{center}
\bigskip 

\monitem
Let $\, f_j(x,t) \, $ be a distribution of particles on the lattice   $\, \cal{ L} \, $ at
the vertex $x$ and discrete time $t$.  
We recall \cite{fhhlpr} (see also  \cite{cd98} or \cite{ll00})  that the discrete dynamics of this 
distribution on the lattice  
$\, \cal{ L} \, $ is given by a collision step followed by a free advection displacement
between two nodes.
We assume that the density 

\moneq
\rho \equiv \sum_j f_j \,
\label{densite}  \end{equation}

\smallskip \noindent
and the momentum 

\moneq
q \equiv \sum_j v_j \, f_j \,
\label{impulsion}  \end{equation}

\smallskip \noindent
 are conserved during the  collision step
and we denote by  $\, f^*_j(x,t) \, $ the distribution after this step: 

\moneq
\rho \equiv  \sum_j f_j =  \sum_j f^*_j  \equiv  \rho^* , \, 
\label{cons-masse}  \end{equation}

\moneq
q  \equiv   \sum_j v_j \, f_j  =  \sum_j v_j \, f^*_j   \equiv   q^* . \, 
\label{cons-impuls}  \end{equation}

\smallskip \noindent 
Then the dynamics of the lattice Boltzmann scheme takes the simple form \cite{du08}

\moneq
f_j (x,\, t +  \Delta t ) \, = \, f^*_j ( x \, - \, v_j \, \Delta t , \,  t)  \, ,
\quad x \in  {\cal{ L}} , \,\,  0 \leq j \leq J \, . \, 
\label{schema}  \end{equation}

\smallskip \noindent
We restrict in what follows to numerical physics that conserve mass and momentum. The
incorporation of conservation of energy is also possible and we refer to 
\cite{ll03}  which discusses various attempts to include energy conservation.
Note that the state of the art concerning the collision step 
$ \, f \longrightarrow f^* \,$  is due to \cite{dh92} with
the so-called ``multiple relaxation time'' Boltzmann scheme. Remark that in all this
contribution the choice of the relaxation model has no influence on our methodology.

\monitem  
{\bf Proposition 1. \quad Conservation property.}

 \noindent We have the following relations concerning the temporal evolution of
  conserved momenta 

\moneq  
\rho(x,\, t + \Delta t)  -  \rho (x,\, t)   +   \sum_j 
\big(    f^*_j (x,\,t)  -  f^*_{\sigma(j)}  ( x_j ,\, t )  \big)  =  0  
\label{masseconserv} \end{equation}

\moneq 
  q(x,\, t + \Delta t)  -  q (x,\, t)     +   \sum_j 
v_j \,   \big( f^*_j (x,\,t)  +  f^*_{\sigma(j)}  ( x_j ,\, t ) \big)   =  0 .
\label{momentconserv} \end{equation}

\smallskip \noindent {\bf Proof of Proposition 1.}

\smallskip \noindent
We have from the dynamics (\ref{schema})  by summation over the index $j$ 

\smallskip \noindent
 $  \displaystyle      \rho(x,\, t + \Delta t) \, \, = \, 
  \sum_j  f^*_j ( x \, - \, v_j \, \Delta t , \,  t)   \, = \,  
\sum_j  f^*_{\sigma(j)}  ( x \, - \, v_{\sigma(j)} \, \Delta t , \, t)  \, = \,   $ 

\noindent $  \displaystyle    \qquad  \qquad  \qquad 
= \sum_j  f^*_{\sigma(j)}  ( x \, + \, v_j  \, \Delta t , \, t)    \, =  \, 
\sum_j  f^*_{\sigma(j)}  ( x_j ,\, t ) $ 

\smallskip \noindent
and the relation (\ref{masseconserv}) is established due to (\ref{cons-masse}). 
In an analogous way, we have for the momentum: 

\smallskip \noindent
$  \displaystyle     q(x,\, t + \Delta t) \,\, = 
  \sum_j  v_j \, f^*_j ( x \, - \, v_j \, \Delta t , \,  t)   \, = \, 
  \sum_j   v_{\sigma(j)} \, f^*_{\sigma(j)} 
( x \, -  \, v_{\sigma(j)} \, \Delta t , \,  t)     \, = \,     $ 

\noindent $  \displaystyle    \qquad  \qquad  \qquad 
=   - \sum_j    v_j \,   f^*_{\sigma(j)}  (  x \, + v_j \, \Delta t , \,  t)      
  =    - \sum_j    v_j \,   f^*_{\sigma(j)}  ( x_j ,\, t )  $ 

\smallskip \noindent
and the relation (\ref{momentconserv}) follows from  (\ref{cons-impuls}).   $\hfill \square$ 

\smallskip   \monitem
We suppose now that we can introduce a cell  $\, K(x) \,$ 
around the vertex $x$ such that its boundary $ \,
\partial K(x) \,$ is composed by $J$ edges $ \, a_j (x)\, $ 
separating the nodes $x$ and $\, x_j \,$: 

\moneq  
 \partial K(x) \,= \, \bigcup_{j > 0} a_j(x) ,\,  
\label{front-K}  \end{equation} 

\smallskip \noindent
with edges $ \, a_j(x) \,$ such that 

\moneq  
 a_j(x) = \partial K(x)  \cap  \partial K(x_j ) = a_{\sigma(j)}(x_j) . \,  
\label{front-arete}  \end{equation} 

\smallskip \noindent
We denote by $ \, \vert K(x)  \vert \,$ and  $ \, \vert  a_j(x)  \vert \,$  the measures of 
 $ \,  K(x)   \,$ and  $ \, a_j(x)  \,$ respectively.
Then the conservation of mass and momentum takes the discrete form 

\moneq 
{{1}\over{\Delta t}} \,  \Big[   \rho (x,\, t+\Delta t)  \,  - \,  
 \rho  (x,\, t) \Big]     
\,+\,  {{1}\over{\vert K(x)  \vert }} \, \sum_j  \vert a_j(x)  \vert \, 
   \psi_j  (x)\, = \, 0 \,,  \label{evol-ro}  \end{equation}

\moneq 
  {{1}\over{\Delta t}} \,  \Big[  q  (x,\, t+\Delta t)  \,  - \,  
 q (x,\, t) \Big]    
 \,+\,  {{1}\over{\vert K(x)  \vert }} \, \sum_j  \vert a_j(x)  \vert \, 
 \zeta_j  (x)\, = \, 0 \,.   \label{evol-q}  \end{equation}

\monitem  
{\bf Proposition 2. \quad   An algebraic expression for general fluxes.}

 \noindent  
We suppose that  the lattice Boltzmann scheme (\ref{cons-masse})  (\ref{cons-impuls})
 (\ref{schema}) satisfies the above hypotheses (\ref{front-K}) and (\ref{front-arete})
and that the control volumes $ \, K(x) \,$ and  $ \, K(x_j) \,$ have the same measure:

\moneq
\vert  K(x) \vert  = \vert  K(x_j) \vert \,, \quad 1 \leq j \leq J \,. 
 \label{egale}  \end{equation}

\smallskip \noindent
We define   the mass flux  $ \,   \psi_j \,$ and the momentum flux $ \,    \zeta_j \,$ 
with the following expressions: 

\moneq
\psi_j(x) \, = \, {{ \vert K(x)  \vert }\over { \Delta t \,  \vert a_j(x)  \vert }} \, 
\big( f^*_j(x) - f^*_{\sigma(j)}(x_j) \big)  \,, \, 
 \label{psi}  \end{equation}

\smallskip 
\moneq
\zeta_j(x) \, = \, {{ \vert K(x)  \vert }\over { \Delta t \,  \vert a_j(x)  \vert }} \, 
v_j \, \big( f^*_j(x) + f^*_{\sigma(j)}(x_j) \big)  \,. 
 \label{zeta} \end{equation}

\smallskip \noindent
Then the quantities defined in  (\ref{psi}) and (\ref{zeta}) are 
so-called   ``conservative fluxes''  in the following sense: 

\moneq 
\psi_j (x) + \psi_{\sigma(j)}(x_j)=0 
 \label{psi-flux} \end{equation}

\moneq 
 \, \zeta_j (x) + \zeta_{\sigma(j)}(x_j)=0 .  \,
 \label{zeta-flux} \end{equation}

\smallskip \noindent
with the vertex  $ \, x_j \,$ defined in  (\ref{sommet}). 

\smallskip  \noindent {\bf Proof of Proposition 2.}

\smallskip \noindent
The first part of the proposition is simply obtained by considering that 
(\ref{masseconserv}) [respectively (\ref{momentconserv})]
and (\ref{evol-ro}) [respectively  (\ref{evol-q})]
  define identically the same evolution equation. Then we have 
for the conservation property of mass: 

\smallskip \noindent  $  \displaystyle   
\psi_j (x) + \psi_{\sigma(j)}(x_j)  =    $ 

\smallskip \noindent $  \displaystyle   \qquad 
= \,   {{ \vert K(x)  \vert }\over { \Delta t \,  \vert a_j(x)  \vert }} \, 
\big( f^*_j(x) - f^*_{\sigma(j)}(x_j)  \big) \,+\, 
 {{ \vert K(x_j)  \vert }\over { \Delta t \,  \vert a_{\sigma(j)}(x_j)  \vert }} \, 
\big( f^*_{\sigma(j)}(x_j) - f^*_{\sigma(\sigma(j))}(x) \big) $ 

\smallskip \noindent $  \displaystyle   \qquad 
= \,   {{ \vert K(x)  \vert }\over { \Delta t \,  \vert a_j(x)  \vert }} \, 
\big( f^*_j(x) - f^*_{\sigma(j)}(x_j)  \big) \,+\, 
 {{ \vert K(x)  \vert }\over { \Delta t \,  \vert a_j(x)  \vert }} \, 
\big( f^*_{\sigma(j)}(x_j) - f^*_{j}(x) \big) $

\smallskip $\hfill$ 
due to  (\ref{front-arete}),  (\ref{egale}) and (\ref{sigma-2}). 

\smallskip \noindent $  \displaystyle   \qquad 
 = 0 .  $ 

\smallskip \noindent
Analogously   for the momentum: 

\smallskip \noindent  
$  \displaystyle  \zeta_j (x) + \zeta_{\sigma(j)}(x_j) \, = \,  
  {{ \vert K(x)  \vert }\over { \Delta t \,  \vert a_j(x)  \vert }} \, 
  v_j \, \big( f^*_j(x) + f^*_{\sigma(j)}(x_j) \big)     \,+\, $

\smallskip \noindent $  \displaystyle   \qquad  \qquad  \qquad  \qquad  \qquad  
  \,+\,   {{ \vert K(x_j)  \vert }\over { \Delta t \,  \vert a_{\sigma(j)}(x_j)  \vert }} \, 
  v_{\sigma(j)} \, \big( f^*_{\sigma(j)}(x_j) + f^*_{\sigma(\sigma(j))}(x)   \big) $ 

\smallskip \noindent $  \displaystyle   \qquad 
= \,   {{ \vert K(x)  \vert }\over { \Delta t \,  \vert a_j(x)  \vert }} \, 
   v_j \, \big( f^*_j(x) + f^*_{\sigma(j)}(x_j) \big)    \,-\, 
 {{ \vert K(x_j)  \vert }\over { \Delta t \,  \vert a_{\sigma(j)}(x_j)  \vert }} \,   
  v_j \, \big( f^*_{\sigma(j)}(x_j) + f^*_j(x)   \big)  $

\smallskip $\hfill$ 
due to  (\ref{symetrie})  and the previous arguments

\smallskip \noindent $  \displaystyle   \qquad 
= \,   0  $

\smallskip \noindent
and the property is established. $ \hfill \square$

\monitem
This remark makes a clear link between the lattice Boltzmann scheme and the finite volume
method \cite{pa80}. Note that the hypothesis (\ref{egale}) can be not satisfied for the
boundary cells as we will see in the following. In that case, we adapt the definition of
the flux in order to enforce the conservation conditions  (\ref{psi-flux})  and 
 (\ref{zeta-flux}).

\section{Flux boundary condition for the  D1Q3 model }\label{d1q3}

\nototo \monitem
In the particular case of  D1Q3  model \cite{qhl}
(see also all algebraic details in \cite{du07}), each vertex $x$ of the lattice has two
neighbours $ \, x_- \equiv x - \Delta x \,$ and  $ \, x_+ \equiv  x + \Delta x . \,$ Then
the number of  particles with velocity equal to $-\lambda$ 
[respectively $0,$ $+\lambda$] is  denoted by 
 $f^-$  [respectively  $f^0$ and  $f^+$]. The bijection $\, \sigma \,$ introduced in 
(\ref{symetrie}) is given simply according to

\moneq 
 \sigma(0) = 0 \,, \quad \sigma(+) = -  \,, \quad \sigma(-) = + \,.
 \label{sym-1D} \end{equation}

\smallskip \noindent 
Moreover, there is a geometrical and topological evidence that a
cell  $\, K(x) \,$ can be constructed around the vertex   $x$:

\moneq 
 K(x) = \big] x - {{\Delta x}\over{2}} \,,\,  x + {{\Delta x}\over{2}} \big[   \,.
 \label{cell-1D} \end{equation}

\smallskip \noindent 
as illustrated in Figure 2. 

\bigskip  
\begin{center}   \includegraphics[height=2cm]     {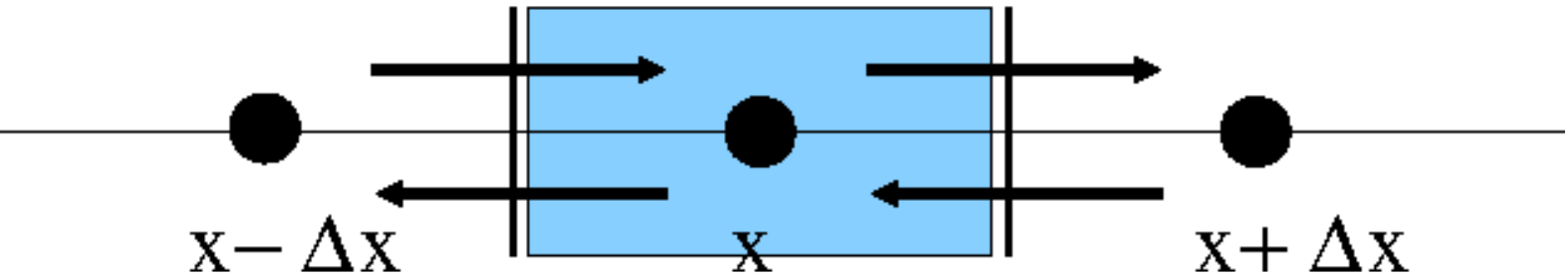} \end{center}
\begin{center} \it Figure 2. Uni-dimensional cell   $\, K(x) \,$ around 
the vertex   $x$. \end{center} \bigskip 

\smallskip \noindent 
We observe that 

\moneq 
  \vert K(x)  \vert  =  \Delta x \,. 
 \label{mes-cell-1D} \end{equation}

\smallskip \noindent 
The  boundary    $ \, \partial K(x) \,$ is composed by    $2$ ``point-like edges''
  $   a_-(x) $ and  $  a_+(x) $  such that  

\moneq 
\vert a_\pm(x)  \vert  =  1 \,.
 \label{front-cell-1D} \end{equation}

\monitem  
{\bf Proposition 3. \quad   Fluxes for the D1Q3 model.}

\noindent
In one space dimension (D1Q3 model) the lattice Boltzmann scheme 
 is exactly a finite volume method.
The mass flux $ \, \psi \, $ and momentum
flux $ \, \zeta \,$ are given by the expressions (\ref {psi}) and (\ref {zeta}) that take
in this particular case the form:  

\moneq 
 \psi_j(x) \, = \, \lambda \,   \big( f^*_j(x) - f^*_{\sigma(j)}(x_j) \big) 
\,, \quad j = - \,,\, 0 \,,\, + \, . 
 \label{psi-1D}  \end{equation}

\moneq  
\zeta_j(x) \, = \, \lambda \,  v_j \,  \big( f^*_j(x) + f^*_{\sigma(j)}(x_j) \big)  
\,, \quad j = - \,,\, 0 \,,\, + \, . 
 \label{zeta-1D}  \end{equation}

\smallskip \noindent {\bf Proof of Proposition 3.}

\smallskip \noindent
We have simply the relations (\ref{masseconserv}) and (\ref{momentconserv}) 
that can be re-written  introducing  (\ref{lambda})  and  (\ref{mes-cell-1D}):

 \smallskip  \noindent $  \displaystyle \qquad 
 {{1}\over{\Delta t}} \, \big(  \rho(x,\, t + \Delta t)  -  \rho (x,\,
 t)  \big)  +   {{1}\over{\Delta x}}   \sum_j  \lambda 
\big(    f^*_j (x,\,t)  -  f^*_{\sigma(j)}  ( x_j ,\, t )  \big)  =  0 \quad $ 

 \smallskip  \noindent $  \displaystyle \qquad 
 \displaystyle {{1}\over{\Delta t}} \, \big(  q(x,\, t + \Delta t)  -  q (x,\,
 t)  \big)  +   {{1}\over{\Delta x}}   \sum_j  \lambda \, v_j \, 
\big(    f^*_j (x,\,t)  +  f^*_{\sigma(j)}  ( x_j ,\, t )  \big)  =  0 \quad $ 

\smallskip \noindent
{\it id est} a vectorial discrete conservation law of the form 

 \smallskip  \noindent $  \displaystyle \qquad 
{{1}\over{\Delta t}} \,  \Big[ W   (x,\, t+\Delta t)  \,  - \, W  (x,\, t) \Big]     \,+\,
{{1}\over{\vert K(x)  \vert }} \, \int_{\partial K}  \Phi  \, \smb \,n \, {\rm d} \gamma
\,  = \, 0 \, \, $

 \smallskip  \noindent 
with a vector $ \, W \, $ composed by density $\, \rho \, $ and momentum $ \, q . \, $ 
Then the relations (\ref{psi-1D}) and  (\ref{zeta-1D}) are clear.  
   $ \hfill \square$

\monitem 
We remark also that we generalize in what follows the terminology ``finite volume
method''. According {\it e.g.} to the classical reference  \cite{gr96}, the definition of
a flux requires {\it a priori}   fluxes to be functions of just the conserved variables. 
Here  mass flux and  momentum flux cannot be expressed in terms of  the only  conserved 
variables (mass and momentum densities) but are in contrary functions of all particle
distributions $f$.

\monitem
We study now the problem of defining a boundary condition for our  D1Q3 Boltzmann model 
\cite{ga94}, \cite{gh03}. We focus on the   particular 
case of the presence of a wall  at one of the extremities.   We suppose that
$x$ is a vertex of the lattice internal to the domain under  study and that its  
right neighbour
$x_+$ is external to the computational domain. Moreover, the geometric position 
$x_{\rm w}$ of the  wall is not supposed to be exactly between $x$ and $x_+$  but at 
a certain fraction $\xi$: 

\moneq  
x_{\rm w} = x + \xi \, \Delta x \,,\qquad 0 < \xi < 1 \, . 
 \label{cl-1D}  \end{equation}

 \smallskip  \noindent 
Note that the particular case $ \, \xi = {{1}\over{2}} \, $ corresponds to a position of
the wall at equal distance between the ``last'' vertex inside the domain and the ``first''
vertex outside the computational domain. 

\bigskip  \bigskip  \setlength\unitlength{1cm}  \begin{picture}(5,3)
\put(1,.5) { \includegraphics[height=2cm]      {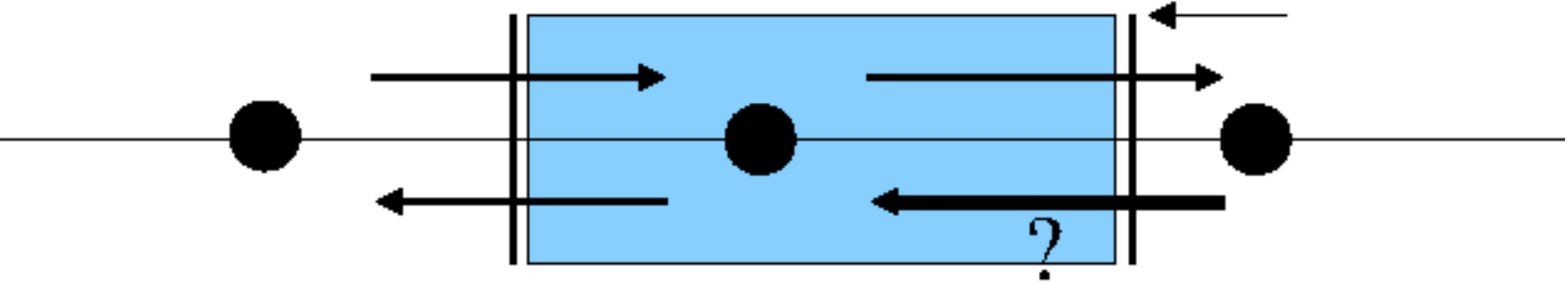} }
\put(4.5,0){$f_-^*(x)$}
\put(4.5,2.8){$f_+^*(x_-)$}
\put(1.8, .8){$x - \Delta x$ }
\put(6.4, .8){$x$ }
\put(10, .8){$x + \Delta x$ }
\put(10.5, 2.3){$x + \xi \Delta x$ }
\put(8,0){$\Phi_-(x)$}
\put(8,2.8){$f_+^*(x)$}
\end{picture} 
\begin{center} \it Figure 3. Numerical boundary condition for the D1Q3 model. 
 \end{center} \bigskip 

\monitem
Assuming  the computational domain has  nontrivial extent, 
we suppose that both vertices $\, x \, $ and  $\, x_- \, $ are 
located inside the computational domain. At a
certain discrete time $\, t ,\, $ we have at our disposal the 
particle transfer $\, f_0^*(x) \,  $  of null velocity at the vertex  $\, x ,\, $ the 
 particle transfer $\, f_-^*(x) \,  $  of speed  $-\lambda$ from vertex $ \, x \, $ to 
the point  $\, x_- , \, $the  particle transfer $ \, f_+^*(x_-) \,  $  of speed $\lambda$ 
from  point $\, x_- \, $  towards vertex  $\, x .\, $ We denote by 
$\, \Phi_-(x) \, $  (instead of  $\, f_-^*(x_+)   $)
the unknown  particle transfer of speed $-\lambda$ coming from the
``ghost'' vertex  $\, x_+   \, $  towards the vertex $x. $    This quantity has to be
determined by the so-called ``numerical boundary scheme''. All the above notations are
illustrated in Figure 3. 

\monitem 
At a boundary vertex $x$, we modify the construction of the control volume $ K(x) $ 
and introduce a natural finite volume defined at the left by the intermediate vertex $x -
{{\Delta x}\over{2}}$ and on the right by the boundary vertex $x_{\rm w}$. Such a control
  volume satisfies

\moneq  
\vert K(x) \vert = \Big( \xi + {1\over2} \Big) \, \Delta x .
\label{mes-bb-cell-1D} \end{equation}

 \smallskip  \noindent 
We observe that $ \vert K(x) \vert$  is equal to $  \Delta x $ only when $ \xi =  {1\over2}
$.

\monitem 
At a boundary, a good numerical methodology is to impose a flux 
(see {\it e.g.} \cite{df89}). This approach  is natural with a finite volume
methodology.  At the solid boundary located at $\, x = x_w ,\,$  
the physical impermeability  condition leads to a  {\bf zero mass flux} 
$ \, \psi_+(x) :\, $    

\moneq  
\psi_+(x) \,=\, 0 \,. 
\label{mass-flux-1D}  \end{equation}

 \smallskip  \noindent 
If we evaluate this mass
flux according to the relation (\ref{psi-1D}), we obtain in this particular case: 
$ \, \psi_+(x) =  \lambda \big( f_+^*(x) - \Phi_-(x) \big) $ and due to 
 (\ref{mass-flux-1D}), we obtain in this manner the so-called ``bounce-back'' boundary
condition: 

\moneq  
\Phi_-(x) = f_+^*(x) \,. 
\label{bb}  \end{equation}

 \smallskip  \noindent 
Other   interpolation schemes have been  proposed by several authors, 
for example   \cite {mls99} \cite {bfl01}.  

\monitem 
{\bf Scheme 1. \quad  Flux  boundary condition  for the D1Q3 model.}

\noindent
Our finite volume boundary condition consists in considering  Figure 3 as a finite
control box $\, K(x) \,$ around vertex $x$ with a particular shape imposed by the geometry
of the problem. The boundary is located at a distance   $  \displaystyle \,  \xi \Delta x$ 
from the vertex $x$. We propose to use the following 
  formula for the unknown input particle number: 

\moneq   
 \Phi_-(x) =  f_+^*(x) + {{\xi -  {1\over2}}\over {\xi +
 {1\over2}}}\,  \Big(    f_-^*(x) - f_+^*(x_-)   \Big) \,.
\label{inputbb-1D} \end{equation}

\smallskip   \noindent {\bf Construction of Scheme 1.}

\smallskip \noindent
We make a   mass balance in a mesh   $K(x)$ 
of measure given by (\ref{mes-bb-cell-1D})  that takes into account the boundary.
In order to enforce the conservation property, the 
 left mass flux  $\, \psi_-(x) \,$ in the direction $x$  $\rightarrow$ $x-\Delta x$ 
is {\it a priori} still 
given according to the relation (\ref{psi-1D})  

\moneq  
\psi_-(x) = \lambda  \big( f_-^*(x) - f_+^*(x-\Delta x) \big) 
\label{psi-bb} \end{equation}

 \smallskip  \noindent 
and the right mass flux $ \,  \psi_+(x) \, $ is null (see (\ref{mass-flux-1D})). 
We then write the time evolution of the scheme in two ways. First, 
we have the general mass
conservation  (\ref{masseconserv})  of a Boltzmann scheme that takes here the form: 

\moneq  
 \rho(x,\, t + \Delta t)  -  \rho (x,\, t)   +  
 \big( f_+^*(x) - \Phi_-(x) \big)  +    \big(  f_-^*(x) - f_+^*(x-\Delta  x)   \big)  
 =  0 \,. 
\label{mascon2} \end{equation}

 \smallskip  \noindent 
Second, we have the mass conservation  (\ref{psi-flux}) inside the volume $\, K(x) $:  

\moneq  
{{1}\over{\Delta t}} \, \big(  \rho(x,\, t + \Delta t)  -  \rho (x,\,
 t)  \big)  +     {{1}\over{ \vert K(x) \vert }}     \Big[  
\psi_+(x) + \psi_-(x)  \Big]   =  0 \,. 
\label{vfibb} \end{equation}

\smallskip \noindent 
We use (\ref{mes-bb-cell-1D}), (\ref{psi-bb}) and the physical condition (\ref{mass-flux-1D})
in order to eliminate   the term 
$ \,  \big(  \rho(x,\, t + \Delta t)  -  \rho (x,\,  t)  \big) \, $ 
between the relations (\ref{mascon2}) and  (\ref{vfibb}). We obtain:

 \smallskip  \noindent $  \displaystyle
   \big(  f_+^*(x) - \Phi_-(x) \big)  + 
  \big(  f_-^*(x) - f_+^*(x-\Delta  x)   \big)  
= {{1}\over{   \xi + {1\over2} }}   \big(  f_-^*(x) - f_+^*(x-\Delta  x)  \big) $ 

 \smallskip  \noindent
and we extract $ \, \Phi_-(x) \,$ from the above expression. Then relation 
 (\ref{inputbb-1D}) is established and the scheme is constructed. $ \hfill \square $

 \monitem 
The numerical boundary scheme  (\ref{inputbb-1D}) has been derived as a consequence of the
mass conservation and a precise treatment of the no-penetration boundary condition 
 (\ref{mass-flux-1D}). This constraint of mass conservation at the boundary has been
studied by \cite{ncgb}. 
Note that with their  own treatment  \cite{gh03}   of the boundary
condition, Ginzburg and D'Humi\`eres have proposed a boundary scheme very close 
to (\ref{inputbb-1D})  that 
conserves  mass in  one space dimension \cite{dh01} \cite{dh06}.

\monitem  
{\bf Proposition 4. \quad  Linearity of the mass flux.} 

 \noindent 
The relation   (\ref{inputbb-1D}) is what is obtained if we suppose that the mass flux
defined in the $x$ direction by the relations

\moneq  
\psi \big(-{{\Delta x}\over{2}} \big) = \lambda  \big(  f_+^*(x-\Delta x) - f_-^*(x)  \big) 
\label{psi-dxs2} \end{equation}

\moneq  
\psi \big({{\Delta x}\over{2}} \big) = \lambda  \big(  f_+^*(x) - \Phi_-(x)  \big) 
\label{psi+dxs2} \end{equation}

\moneq  
\psi \big(\xi \, \Delta x \big) = 0 
\label{psibord} \end{equation}

 \noindent
at respective positions $-{{\Delta x}\over{2}}$,  ${{\Delta x}\over{2}}$ and 
$\xi \, \Delta x$ at the boundary is {\bf linear}.

\smallskip \noindent {\bf Proof of Proposition 4.}

\smallskip \noindent We remark first that $ \psi (-{{\Delta x}\over{2}}) = -  \psi_-(x) $
(defined in  (\ref{psi-bb}))  due to the choice of the direction to measure this mass
flux. The condition of linearity for the function defined by the relations
(\ref{psi-dxs2}), (\ref{psi+dxs2}) and (\ref{psibord}) can be expressed under the form 

 \smallskip \qquad  $  \displaystyle
{{0 -  \lambda  \big(  f_+^*(x-\Delta x) - f_-^*(x)  \big) }\over{ \xi \, \Delta x - 
( -{{\Delta x}\over{2}}) }} = {{ 0 -  \lambda  \big(  f_+^*(x) - \Phi_-(x)  \big) }\over
{  \xi \, \Delta x -  {{\Delta x}\over{2}} }}  $ 

 \smallskip  \noindent
that is 

 \smallskip   \qquad $  \displaystyle
{{   f_+^*(x-\Delta x) - f_-^*(x)   }\over{ \xi + {{1}\over{2}} }} = 
{{   f_+^*(x) - \Phi_-(x) }\over {  \xi -  {{1}\over{2}} }}  $ 

 \smallskip  \noindent
and the last relation corresponds exactly to   (\ref{inputbb-1D}).    $ \hfill \square$

\bigskip \bigskip

\monitem 
{\bf  Uni-dimensional acoustic wave.}

\bigskip  
\begin{center}   \includegraphics[height=2.5cm]     {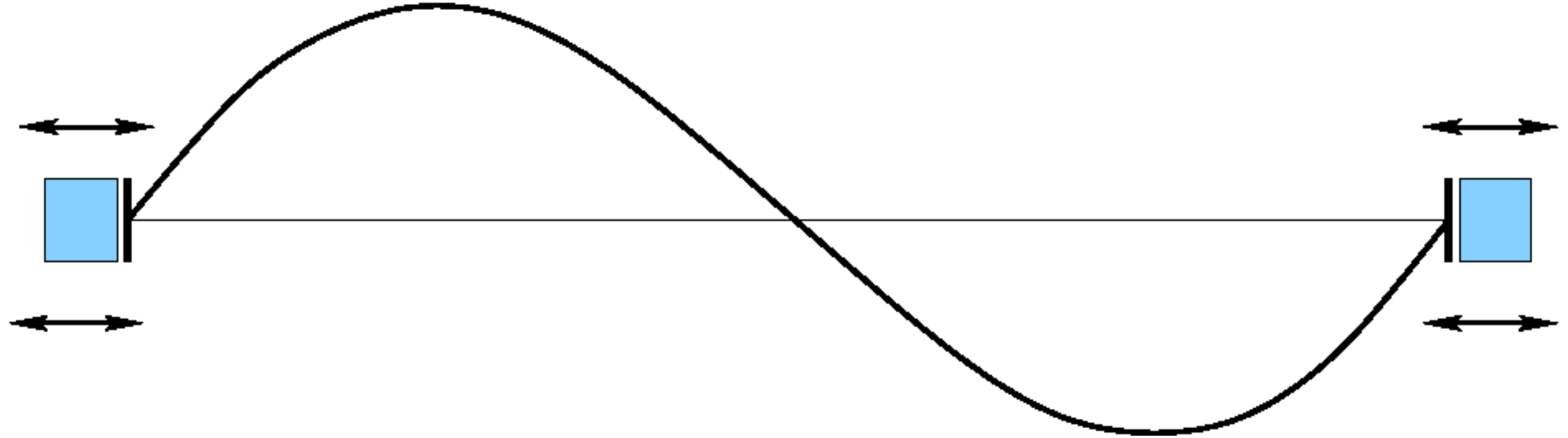} \end{center}
\begin{center} \it Figure 4. Uni-dimensional acoustic wave. 
 \end{center} \bigskip 

\begin{center}   \includegraphics  [height=8.5cm] 
{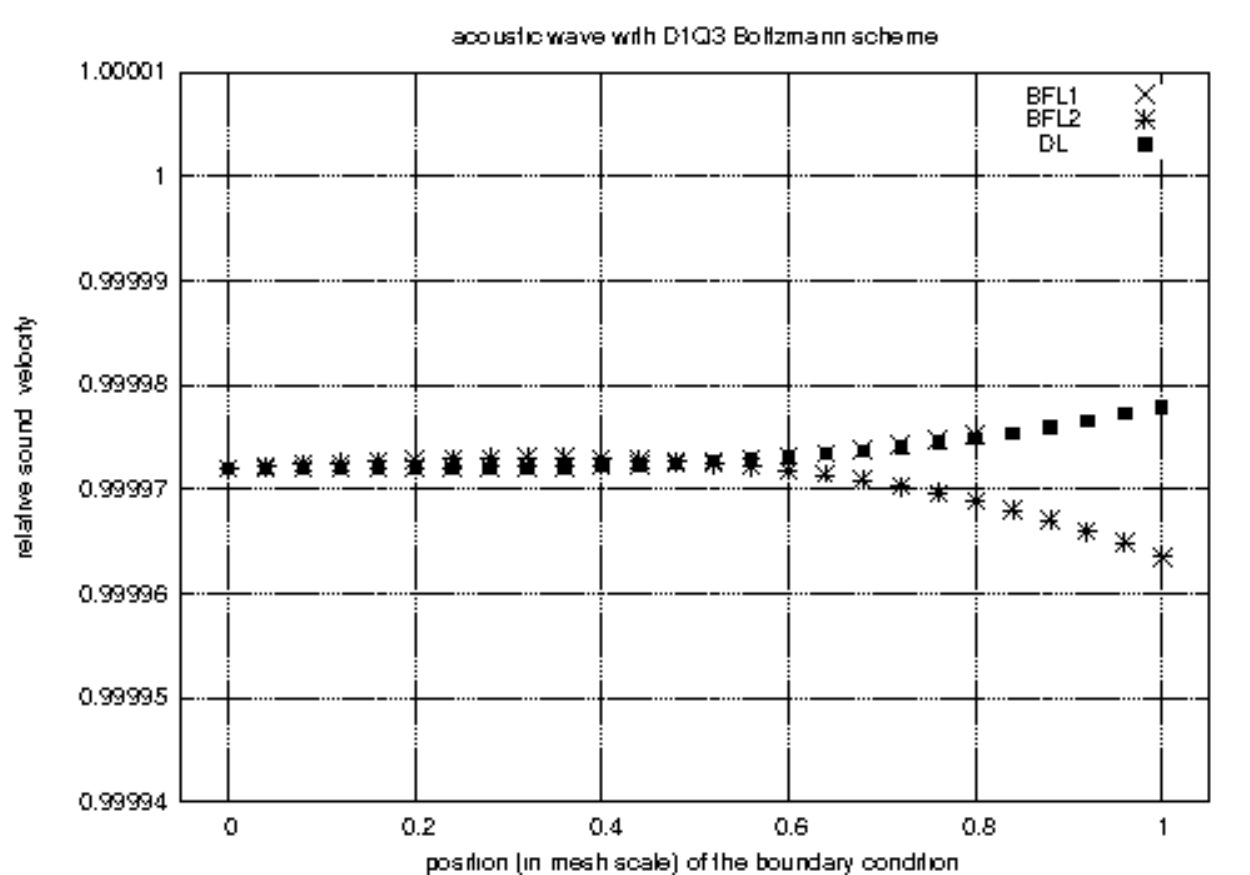} \end{center}
\begin{center} \it Figure 5.  Relative sound velocity with D1Q3 lattice Boltzmann scheme for
  various schemes and a mesh of 100 points.  \end{center}   
\bigskip 

\monitem
We have tested the above idea in the case of an acoustic wave in a tube closed at the two
extremities. We are able to produce a 
variation of the boundary condition location by a fraction  $ \, \xi \, $  of the mesh 
$ \Delta x $ (see Figure 4). 
We determine the eigenvalues of the operator that corresponds to one time step of the
LB algorithm, using the ARPACK \cite{arpack} software package. For the lowest mode
with effective wave vector 

\moneq    
k = {{\pi}\over{(N-1 +2 \xi) \Delta x}}
\label{vecteurond} \end{equation}

 \smallskip  \noindent
 this leads to 
$ \, -\gamma_r + i \gamma_i \, $ from which we determine an effective speed of sound

\moneq  
c_{\rm rel}= {{   \gamma_i }\over { c_s \, k }}  
\label{cson} \end{equation}

 \smallskip  \noindent
if there are $N$ lattice points between the boundaries. We introduce similarly the
effective relative attenuation 

\moneq  
a_{\rm rel}= {{   \gamma_r }\over { {1\over2}  \,  \nu_{\ell}  \, k^2 }}
\label{att} \end{equation}

 \smallskip  \noindent
with $\, \nu_{\ell} \,$ the longitudinal kinematic viscosity  \cite{ll59}. 
The relative value of sound velocity  is displayed in Figure 5 and
Table 1. 
The results of our scheme are comparable with those of  Bouzidi {\it al.} 
\cite{bfl01} when using linear extrapolation. 
After a simple exploitation of Table 1 with least squares, 
the error for sound velocity with 
bounce-back scheme is proportional to  $ \, {{1}\over{N}}  \,$    whereas
it is proportional to  $ \, {{1}\over{N^2}}  \,$ for 
 both  versions of the Bouzidi scheme and our scheme. 
 
\bigskip \bigskip \noindent 

\setbox20=\hbox{ $\,\,$ }
\setbox30=\hbox{ 100 points }
\setbox50=\hbox{ 200 points }
\setbox60=\hbox{ 300 points }
\setbox21=\hbox{ bounce-back }
\setbox31=\hbox{ $ 9.97231  \times 10^{-3}$ }
\setbox51=\hbox{ $ 4.99309  \times 10^{-3}$ }
\setbox61=\hbox{ $ 3.33025  \times 10^{-3}$ }
\setbox22=\hbox{ BFL1 }
\setbox32=\hbox{ $ 2.797 \times 10^{-5}$  } 
\setbox52=\hbox{ $ 6.94  \times 10^{-6}$  }
\setbox62=\hbox{ $ 3.09  \times 10^{-6}$  }
\setbox23=\hbox{ BFL2 }
\setbox33=\hbox{ $ 3.645 \times 10^{-5} $  }
\setbox53=\hbox{ $ 8.02  \times 10^{-6} $  }
\setbox63=\hbox{ $ 3.41  \times 10^{-6} $  }
\setbox24=\hbox{ DL }
\setbox34=\hbox{ $ 2.803   \times 10^{-5}$  }
\setbox54=\hbox{ $ 6.95    \times 10^{-6}$ }
\setbox64=\hbox{ $ 3.09   \times 10^{-6}$}
\setbox44=\vbox{\offinterlineskip  \halign {
&\tvg#& # &\tvg#&   #  &\tvg#&  #  &\tvg#&  # &\tvd#\cr 
\na&  \box20 && \box30 &&  \box50  && \box60  \hcr 
\na&  \box21 && \box31 &&  \box51  && \box61  \hcr 
\na&  \box22 && \box32 &&  \box52  && \box62  \hcr 
\na&  \box23 && \box33 &&  \box53  && \box63  \hcr 
\na&  \box24 && \box34 &&  \box54  && \box64  \hcr 
 \na}   }  \centerline{\box44  }

\begin{center} \it Table 1. Largest discrepancy of the relative 
sound velocity with D1Q3 lattice Boltzmann scheme for
  various boundary schemes and meshes. \end{center}

 \bigskip 

\monitem
We give in Figure 6 and Table 2 various results for the effective attenuation 
$ a_{\rm rel} .$ 
Our method is spectacularly better than the linear extrapolation case (``BFL1'') and
comparable with the quadratic interpolation scheme  (``BFL2'') of the previous authors  
\cite{bfl01} when the boundary is not located
exactly half-way  between two mesh
points and equivalent to the previous one in this particular geometric case.
After an elementary exploitation of Table 2, bounce-back and linear extrapolation version
of Bouzidi scheme give an error for attenuation of the first eigenmode proportional to 
$ \, {{1}\over{N}} . \,$ This error for attenuation is  proportional to 
$ \, {{1}\over{N^2}}  \,$ with quadratic  extrapolation version of Bouzidi scheme
and  proportional to  $ \, {{1}\over{N^3}}  \,$ with the present scheme.


\begin{center}   \includegraphics[height=8.5cm]  
{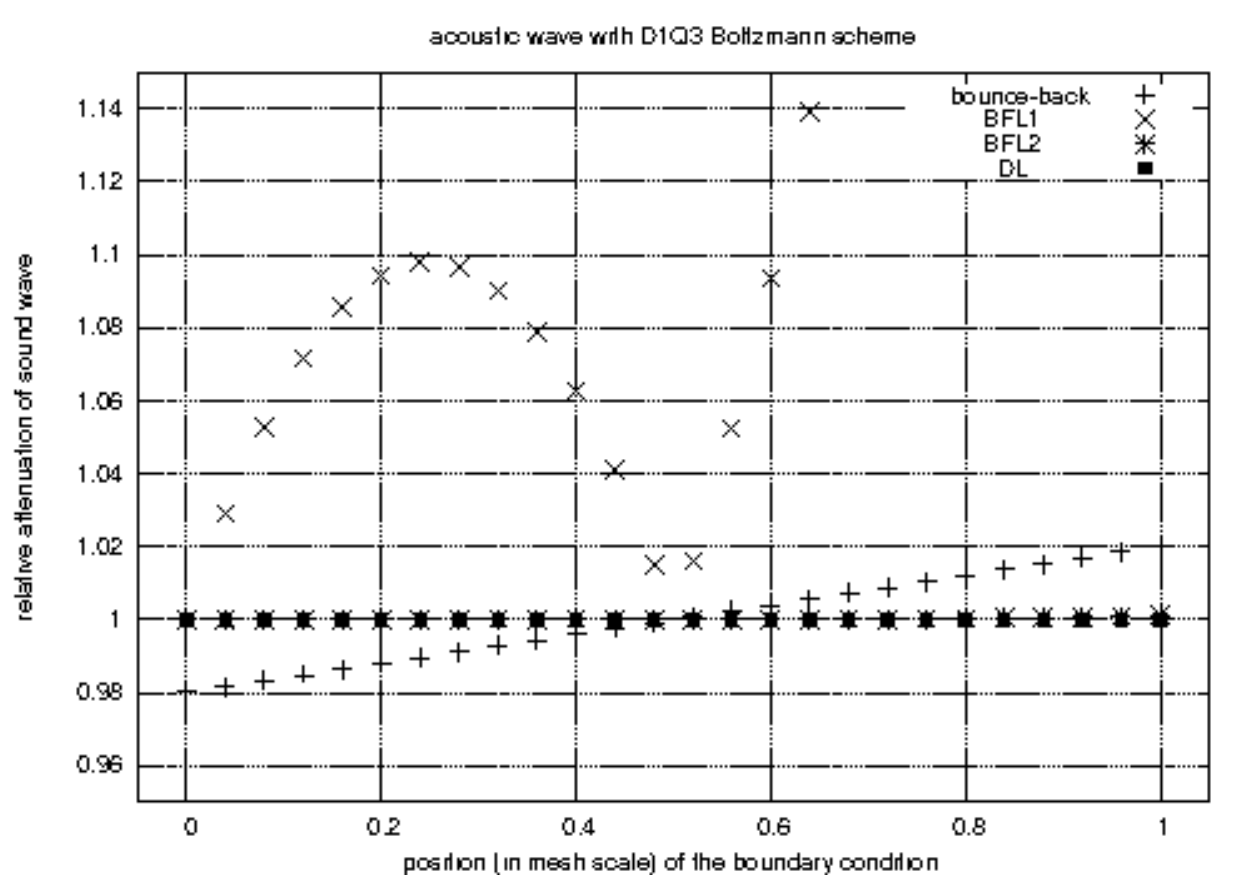} \end{center}
\begin{center} \it Figure 6.   Relative attenuation of an acoustic  wave for various
numerical boundary Boltzmann schemes and a mesh of 100 points.    \end{center}
\bigskip 


\bigskip \noindent 

\setbox20=\hbox{ $\,\,$ }
\setbox30=\hbox{ 100 points }
\setbox50=\hbox{ 200 points }
\setbox60=\hbox{ 300 points }
\setbox21=\hbox{ bounce-back }
\setbox31=\hbox{ $ 2.009999  \times 10^{-2} $  }
\setbox51=\hbox{ $ 1.002496  \times 10^{-2} $  }
\setbox61=\hbox{ $ 6.67772   \times 10^{-3} $  } 
\setbox22=\hbox{ BFL1 }
\setbox32=\hbox{ 0.77150864 }
\setbox52=\hbox{ 0.38796054 }
\setbox62=\hbox{ 0.25910648 }
\setbox23=\hbox{ BFL2 }
\setbox33=\hbox{ $ 1.10138 \times 10^{-3} $ }
\setbox53=\hbox{ $ 1.3972  \times 10^{-4} $ }
\setbox63=\hbox{ $ 4.157  \times 10^{-5} $ }
\setbox24=\hbox{ DL }
\setbox34=\hbox{ $ 9.57 \times 10^{-6} $  }
\setbox54=\hbox{ $ 1.17  \times 10^{-6} $ }
\setbox64=\hbox{ $ 3.1 \times 10^{-7} $  }
\setbox44=\vbox{\offinterlineskip  \halign {
&\tvg#& # &\tvg#&   #  &\tvg#&  #  &\tvg#&  # &\tvd#\cr 
\na&  \box20 && \box30 &&  \box50  && \box60  \hcr 
\na&  \box21 && \box31 &&  \box51  && \box61  \hcr 
\na&  \box22 && \box32 &&  \box52  && \box62  \hcr 
\na&  \box23 && \box33 &&  \box53  && \box63  \hcr 
\na&  \box24 && \box34 &&  \box54  && \box64  \hcr 
 \na}   }  \centerline{\box44  }

\begin{center} \it Table 2.   Largest discrepancy of the relative 
attenuation of an acoustic  wave for various
numerical boundary Boltzmann schemes and meshes.   \end{center}

 



\section{Finite volumes for the  D2Q9 model} \label{d2q9}

\nototo \monitem
We have two formulae for the time evolution of the conserved momenta: 
the evolution of mass (\ref{masseconserv}) and momentum (\ref{momentconserv}) that comes
from the general properties of a lattice Boltzmann scheme and the finite volumes  
 framework (\ref{evol-ro})  (\ref{evol-q}). For the case of the two dimensional
model D2Q9 (defined {\it e.g.} in   \cite{qhl}), we set two natural questions: 
{\it  (i)}   Where is  (geometrically~!!) the  finite volume  $\,  K(x)  \,$ ? 
{\it  (ii)}  What are the possible formulae for the mass flux $\,  \psi_j \,$ 
and the momentum flux   $\,  \zeta_j \,$  ?  To  our knowledge, there is 
no satisfying answer  to the above questions ! 
We suggest here to use  {\bf two}   different control finite volumes 
$ \,  K_{\|} \, $ and   $ \,  K_{\times } \,$ defined in Figure~7
and essentially to neglect the internal dynamics between the two control volumes during
the relaxation   step.

\bigskip  
\begin{center}   \includegraphics[height=6.5cm]   {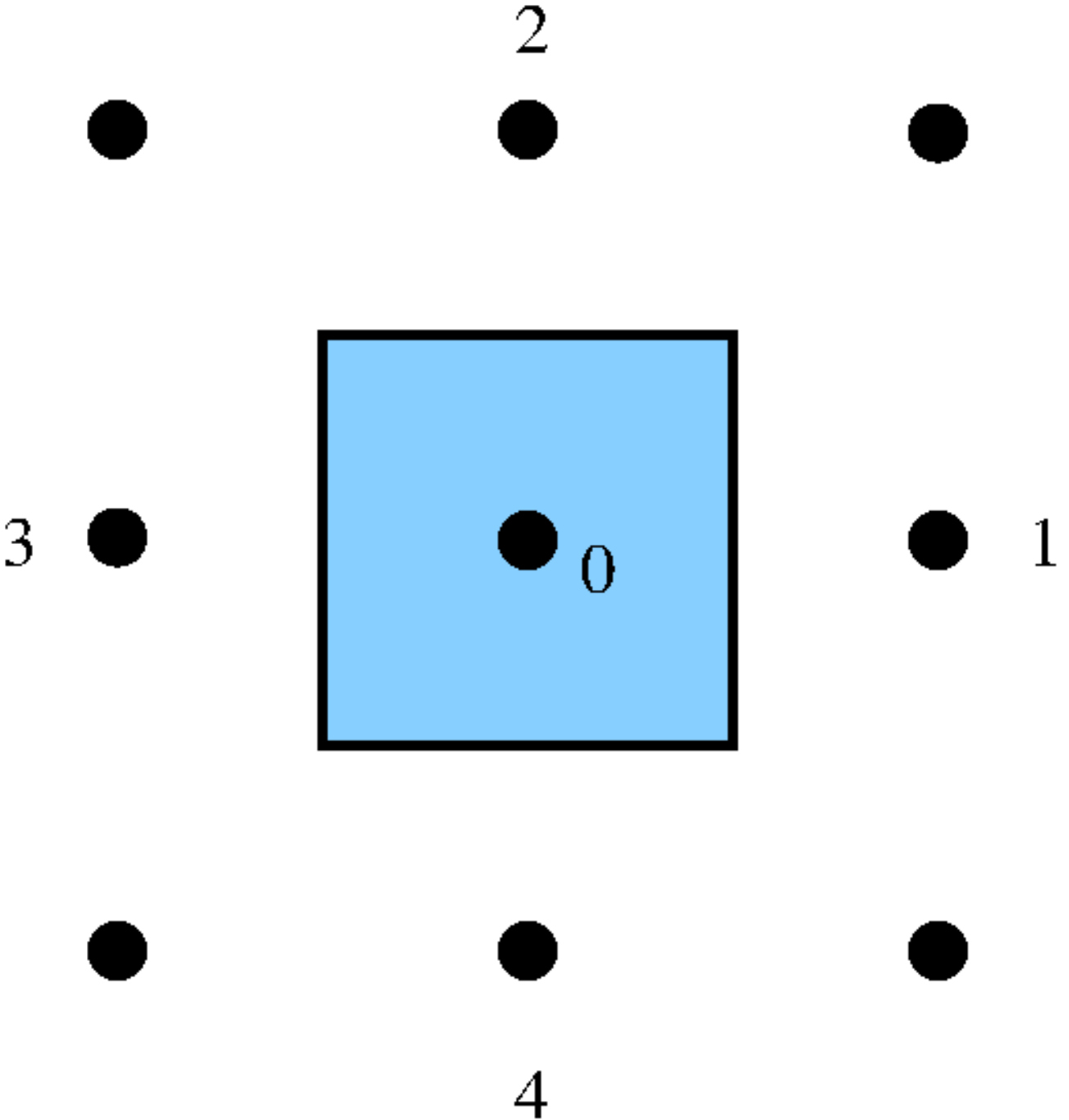} 
\qquad \qquad   \includegraphics[height=6.5cm]    {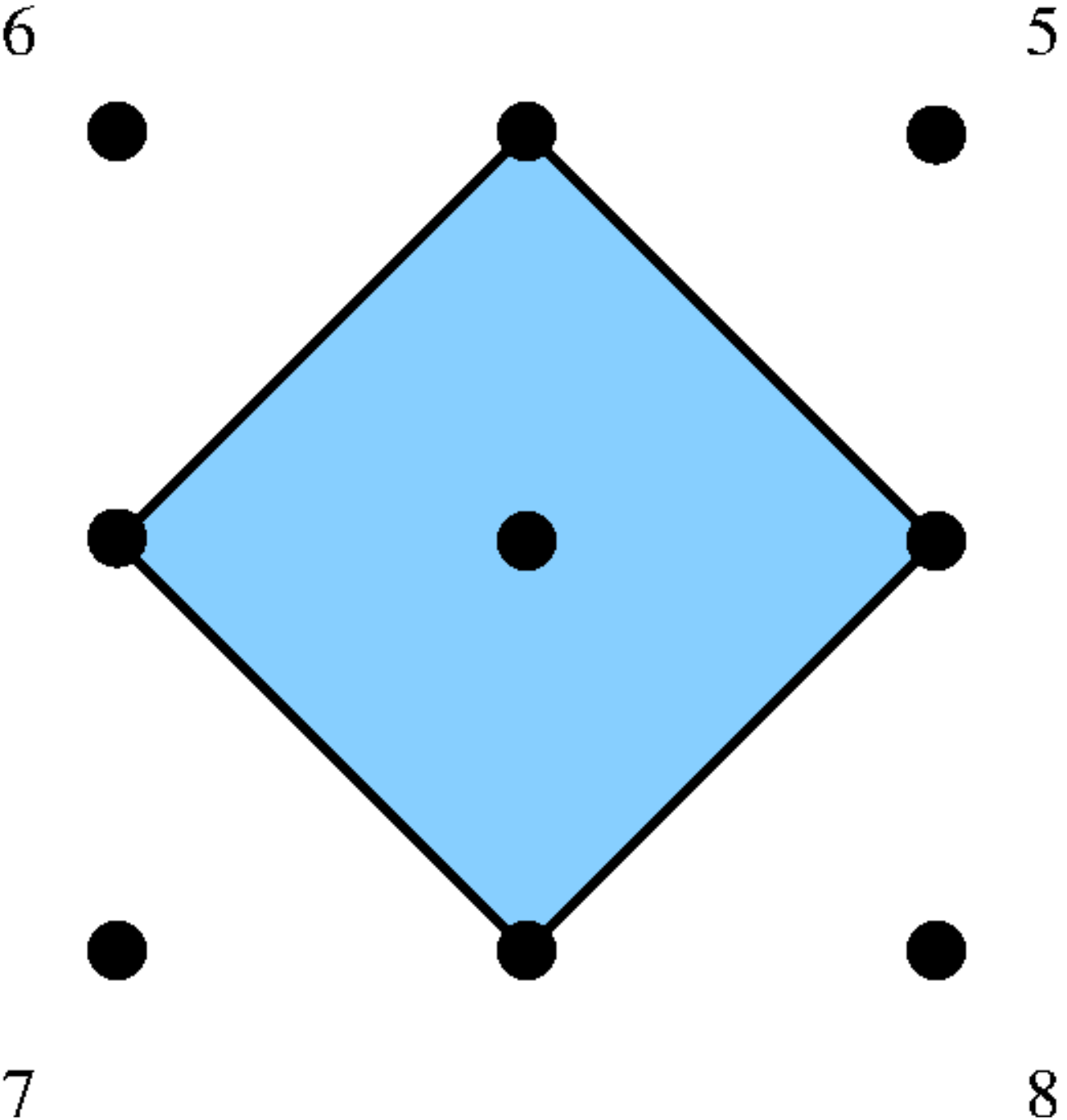}  \end{center}
\begin{center} \it Figure 7. Control finite  volumes $ \, K _{\|} \,$  and   
$ \, K _{\times} \,$  for the two-dimensional D2Q9 lattice Boltzmann scheme.
 \end{center} \bigskip 

 \monitem
We look carefully at  Figure 7 and we observe that 

\moneq   
 \vert  K_{\|} \vert = \Delta x ^2 \,, \qquad  
\vert  K_{\times } \vert = 2 \, \Delta x ^2 \,. 
\label{vol-partiel} \end{equation}

 \smallskip  \noindent
Moreover the boundary $\, \partial  K_{\|} \,$ [respectively  $\, \partial  K_{\times} $]
is composed by the four edges $\, a_j \,$
for $j = 1$ to $ 4 $   [respectively  $j = 5$ to $ 8 $] and we have

\moneq   
 \vert  a_j  \vert = \Delta x  \,, \,\, j = 1, \, 2 ,\, 3 ,\, 4 \,, \qquad 
 \vert  a_j  \vert = \sqrt{2} \, \Delta x  \,, \,\, j = 5, \, 6 ,\, 7 ,\, 8 \,.
\label{mes-edges} \end{equation}

 \monitem
We introduce the partial densities 
 $\, \rho_{\|} (x,\, t) ,\,$  $\, \rho_{\times} (x,\, t) \,$ 
and the partial momenta  $\, q_{\|} (x,\, t) ,\,$  $\, q_{\times} (x,\, t) \,$
according to 

\moneq   
 \rho_{\|} (x,\, t) =    \sum_{j=0}^4  f_j (x,\,t) \,, \qquad 
 \rho_{\times} (x,\, t) =    \sum_{j=5}^8  f_j (x,\,t) \,
\label{rho-partiel} \end{equation}

\moneq  
   q_{\|} (x,\, t) =    \sum_{j=0}^4  v_j \, f_j (x,\,t) \,, \qquad 
  q_{\times} (x,\, t) =    \sum_{j=5}^8  v_j \, f_j (x,\,t) \,
\label{q-partiel} \end{equation}

\smallskip \noindent 
and the analogous quantities  
$\, \rho_{\|}^* (x,\, t) ,\,$  $\, \rho_{\times}^* (x,\, t) \,$ 
$\, q_{\|}^* (x,\, t) ,\,$  $\, q_{\times}^* (x,\, t) \,$
by replacing $ \, f \,$ by $\, f^* \,$ after collisions in the relations 
(\ref{rho-partiel}) and (\ref{q-partiel}).
We introduce also the defect of conservation of the partial momenta: 

\moneq   
\Delta \rho \equiv  \rho_{\|}^* (x,\, t) -  \rho_{\|} (x,\, t)   \,, \qquad  
\Delta q \equiv   q_{\|}^* (x,\, t) -  q_{\|} (x,\, t) \, . 
\label{deltas} \end{equation}

\monitem  
{\bf Proposition 5. \quad  Internal defect of conservation.}

 \noindent  
With the above definitions, we have 

\moneq    
 \rho_{\|}(x,\, t + \Delta t)  -  \rho_{\|} (x,\,  t)   
+   \sum_{j=0}^4     \big(    f^*_j (x,\,t)  -  f^*_{\sigma(j)}  ( x_j ,\, t )  \big) 
= \Delta \rho 
\label{part-rho-par} \end{equation}

\moneq    
 \rho_{\times}(x,\, t + \Delta t)  -  \rho_{\times}(x,\, t )  
+   \sum_{j=5}^8     \big(    f^*_j (x,\,t)  -  f^*_{\sigma(j)}  ( x_j ,\, t )  \big) 
= - \Delta \rho 
\label{part-rho-croix} \end{equation}

\moneq    
 q_{\|}(x,\, t + \Delta t)  - q_{\|} (x,\,  t)  
+   \sum_{j=0}^4    \, v_j \,   \big(    f^*_j (x,\,t)  +  f^*_{\sigma(j)}  ( x_j ,\, t )  \big) 
= \Delta q
\label{part-q-par} \end{equation}

\moneq    
 q_{\times}(x,\, t + \Delta t)  - q_{\times} (x,\,  t)  
+   \sum_{j=5}^8    \, v_j \,   \big(    f^*_j (x,\,t)  +  f^*_{\sigma(j)}  ( x_j ,\, t )  \big) 
= -\Delta q \,. 
\label{part-q-croix} \end{equation}

\smallskip \noindent {\bf Proof of Proposition 5.}

\smallskip \noindent
It is a direct consequence of the definitions 
(\ref{rho-partiel}), (\ref{q-partiel}), (\ref{deltas}) and of the microscopic iteration of
the scheme (\ref{schema}). To fix the ideas, we detail the proof of 
 (\ref{part-rho-par}): 

 \smallskip  \noindent $  \displaystyle
 \rho_{\|}(x,\, t + \Delta t)  -  \rho_{\|} (x,\,  t) = 
  \sum_{j=0}^4   f^*_j (x - v_j \Delta t ,\,t) -  \rho_{\|}^* (x,\,  t) + \Delta \rho $

 \smallskip  \noindent $  \displaystyle \qquad  \qquad 
=  \sum_{j=0}^4   f^*_{\sigma(j)} (x - v_j \Delta t,\,t) -  
\sum_{j=0}^4   f^*_j (x ,\,t)   + \Delta \rho $

 \smallskip  \noindent $  \displaystyle \qquad  \qquad 
=  -  \sum_{j=0}^4 \big(  f^*_j (x ,\,t)  - f^*_{\sigma(j)} (x - v_j \Delta t,\,t) 
\big)   + \Delta \rho \, . $             $ \hfill \square$ 

 \monitem
In what follows, we neglect the difference between 
$ \,  \rho_{\|}(x,\, t + \Delta t)  -  \rho_{\|} (x,\,  t)  \,$ 
and 
$ \,  \rho_{\|}(x,\, t + \Delta t)  -  \rho_{\|}^* (x,\,  t)  \,$ 
when   we suppose  that $\, \Delta \rho \,$ is equal to zero. 
In other terms, the partial masses $\,  \rho_{\|} \,$ and 
 $\,  \rho_{\times} \,$ are supposed to be conserved 
during the collision process.  Of course, we make the same
hypothesis for the momentum and the differences  
$ \,  q_{\|}(x,\, t + \Delta t)  -  q_{\|} (x,\,  t)  \,$ 
and 
$ \,  q_{\|}(x,\, t + \Delta t)  -  q_{\|}^* (x,\,  t)  \,$ 
are neglected when   $\, \Delta q \,$ is supposed to be negligeable. 
We have the following proposition that uses explicitly the 
less natural  increment 
$ \,  \rho_{\|}(x,\, t + \Delta t)  -  \rho_{\|}^* (x,\,  t)  \,$ 
and associated.

\monitem  
{\bf Proposition 6. \quad  Partial numerical fluxes.} 

 \noindent 
We have the following expressions for the
time evolution 

\moneq 
{{1}\over{\Delta t}} \,  \Big[   \rho_{\|} (x,\, t+\Delta t)  \,  - \,  
 \rho_{\|}^*  (x,\, t) \Big]     
\,+\,  {{1}\over{\vert K_{\|}  \vert }} \, \sum_{j=0}^4  \vert a_j  \vert \, 
   \psi_j  (x)\, = \, 0 \,,  \label{evol-ro-paral}  \end{equation}

\moneq 
{{1}\over{\Delta t}} \,  \Big[   \rho_{\times} (x,\, t+\Delta t)  \,  - \,  
 \rho_{\times}^*  (x,\, t) \Big]     
\,+\,  {{1}\over{\vert K_{\times}  \vert }} \, \sum_{j=5}^8  \vert a_j  \vert \, 
   \psi_j  (x)\, = \, 0 \,,  \label{evol-ro-croix}  \end{equation}

\moneq 
{{1}\over{\Delta t}} \,  \Big[   q_{\|} (x,\, t+\Delta t)  \,  - \,  
q_{\|}^*  (x,\, t) \Big]     
\,+\,  {{1}\over{\vert K_{\|}  \vert }} \, \sum_{j=0}^4  \vert a_j  \vert \, 
   \zeta_j  (x)\, = \, 0 \,,  \label{evol-q-paral}  \end{equation}

\moneq 
{{1}\over{\Delta t}} \,  \Big[  q_{\times} (x,\, t+\Delta t)  \,  - \,  
q_{\times}^*  (x,\, t) \Big]     
\,+\,  {{1}\over{\vert K_{\times}  \vert }} \, \sum_{j=5}^8  \vert a_j  \vert \, 
   \zeta_j  (x)\, = \, 0 \,,  \label{evol-q-croix}  \end{equation}

\smallskip \noindent
with  ``mass fluxes''  $\,  \psi_j (x)  \,$ given by

\moneq 
\psi_j  (x) =   \lambda \, 
\big(    f^*_j (x,\,t)  -  f^*_{\sigma(j)}  ( x_j ,\, t )  \big) \,, 
\,\,\, \qquad 0  \leq j \leq 4 \,
\label{psi-para}  \end{equation}

\moneq 
 \psi_j  (x) =   \lambda \, \sqrt{2} \, 
\big(    f^*_j (x,\,t)  -  f^*_{\sigma(j)}  ( x_j ,\, t )  \big) \,, 
\quad 5 \leq j \leq 8 \, 
\label{psi-croix}  \end{equation}

\smallskip \noindent
and ``momentum fluxes''  $\,  \zeta_j (x)  \,$  by

\moneq 
 \zeta_j  (x) =   \lambda \, v_j \, 
\big(    f^*_j (x,\,t)  +  f^*_{\sigma(j)}  ( x_j ,\, t )  \big) \,,
 \, \, \,  \qquad 0 \leq j \leq 4 \,
\label{zeta-para}  \end{equation}

\moneq 
 \zeta_j  =   \lambda \, v_j \, \, \sqrt{2} \, 
\big(    f^*_j (x,\,t)  +  f^*_{\sigma(j)}  ( x_j ,\, t )  \big) \,, 
\quad 5 \leq j \leq 8 \,. 
\label{zeta-croix}  \end{equation}

\smallskip \noindent {\bf Proof of Proposition 6.}

\smallskip \noindent
We simply make the partial sums from the fundamental evolution relation   
(\ref{schema}) and we get by introducing the space scale $\, \Delta x ,\,$ the time scale 
 $\, \Delta t ,\,$ and their ratio $\, \lambda : \,$ 

 \smallskip  \noindent $  \displaystyle
{{1}\over{\Delta t}} \, \big(  \rho_{\|}(x,\, t + \Delta t)  -
 \rho_{\|}^* (x,\,  t)  \big)  +   {{1}\over{\Delta x}}  \sum_{j=0}^4     \lambda 
\big(    f^*_j (x,\,t)  -  f^*_{\sigma(j)}  ( x_j ,\, t )  \big)  =  0 \quad $ 

 \smallskip  \noindent $  \displaystyle
{{1}\over{\Delta t}} \, \big(  \rho_{\times}(x,\, t + \Delta t)  -
 \rho_{\times}^* (x,\,  t)  \big)  +   {{1}\over{\Delta x}}   \sum_{j=5}^8    \lambda 
\big(    f^*_j (x,\,t)  -  f^*_{\sigma(j)}  ( x_j ,\, t )  \big)  =  0 \quad $ 

 \smallskip  \noindent $  \displaystyle
 {{1}\over{\Delta t}} \, \big(  q_{\|}(x,\, t + \Delta t)  -  q_{\|}^* (x,\,
 t)  \big)  +   {{1}\over{\Delta x}}   \sum_{j=0}^4  \lambda \, v_j \, 
\big(    f^*_j (x,\,t)  +  f^*_{\sigma(j)}  ( x_j ,\, t )  \big)  =  0 \quad $ 

 \smallskip  \noindent $  \displaystyle
 {{1}\over{\Delta t}} \, \big(  q_{\times}(x,\, t + \Delta t) 
 -  q_{\times}^* (x,\,
 t)  \big)  +   {{1}\over{\Delta x}}   \sum_{j=5}^8  \lambda \, v_j \, 
\big(    f^*_j (x,\,t)  +  f^*_{\sigma(j)}  ( x_j ,\, t )  \big)  =  0 \, .  $ 

 \smallskip  \noindent 
We replace in the   above expressions the  space scale  $\, \Delta x ,\,$
by the correct expression as function of (\ref{vol-partiel}) and (\ref{mes-edges}):

 \smallskip  \noindent $  \displaystyle
{{1}\over{\Delta t}} \, \big(  \rho_{\|}(x,\, t + \Delta t)  -
 \rho_{\|}^* (x,\,  t)  \big)  +  
 \sum_{j=0}^4    { { \vert  a_j \vert  } \over { \vert K _{\|}   \vert   }}    
   \, \lambda \, 
\big(    f^*_j (x,\,t)  -  f^*_{\sigma(j)}  ( x_j ,\, t )  \big)  =  0 \quad $ 

 \smallskip  \noindent $  \displaystyle
{{1}\over{\Delta t}} \, \big(  \rho_{\times}(x,\, t + \Delta t)  -
 \rho_{\times}^* (x,\,  t)  \big)   + 
\sum_{j=5}^8 { { \vert  a_j \vert  } \over { \vert K _{\times}   \vert   }}  
\, \lambda \, \sqrt{2}  
\big(    f^*_j (x,\,t)  -  f^*_{\sigma(j)}  ( x_j ,\, t )  \big)  =  0 \quad $ 

 \smallskip  \noindent $  \displaystyle
{{1}\over{\Delta t}} \, \big(  q_{\|}(x,\, t + \Delta t)  -  q_{\|}^* (x,\,
 t)  \big)    +  
 \sum_{j=0}^4    { { \vert  a_j \vert  } \over { \vert K _{\|}   \vert   }}    
 \, \lambda \,   v_j \, 
\big(    f^*_j (x,\,t)  +  f^*_{\sigma(j)}  ( x_j ,\, t )  \big)  =  0 \quad $ 

 \smallskip  \noindent $  \displaystyle
{{1}\over{\Delta t}} \, \big(  q_{\times}(x,\, t + \Delta t)  
-  q_{\times}^* (x,\,  t)  \big)    +  
\sum_{j=5}^8 { { \vert  a_j \vert  } \over { \vert K _{\times}   \vert   }}    
\, \lambda \, \sqrt{2} \,  v_j \, 
\big(    f^*_j (x,\,t)  +  f^*_{\sigma(j)}  ( x_j ,\, t )  \big)  =  0  \, . $ 

 \smallskip  \noindent 
We have clearly an exact equivalence between the four  above  expressions 
with the relations 
(\ref{evol-ro-paral}), (\ref{evol-ro-croix}), (\ref{evol-q-paral})
and  (\ref{evol-q-croix}) if we make the choices 
(\ref{psi-para}), (\ref{psi-croix}), (\ref{zeta-para}) and  (\ref{zeta-croix})
 for the fluxes.   $ \hfill \square$ 

\monitem
We remark that we have proposed to cut the density (and the momentum) into two parts 
$\,  \rho_{\|} \,$ and $\,  \rho_{\times} : \,$ 
 $ \, \rho \equiv  \rho_{\|} +  \rho_{\times} \, $
and we have also  $ \,  \rho^* \equiv  \rho_{\|}^* +  \rho_{\times}^* \, $
but keep in  memory that we do not have a conservation law for the partial densities:
$  \,   \rho_{\|}^* \not=  \rho_{\|} \, $ and $  \,   \rho_{\times}^* \not=  
\rho_{\times} \, $ 
{\it a priori} even if $  \,   \rho^* \equiv  \rho .  \, $ 
We have a similar remark for the momentum: 
$  \,   q_{\|}^* \not=  q_{\|} \, $ and $  \, q_{\times}^* \not= q_{\times} \, $ 
{\it a priori}.  Therefore the relations (\ref{evol-ro-paral}), 
 (\ref{evol-ro-croix}),  (\ref{evol-q-paral}) and  (\ref{evol-q-croix}) are 
algebraically exact but are  not a rigorous
discretization of the conservation laws of mass and momentum. They have to be  seen 
as a first tentative to merge a Boltzmann scheme inside the finite volume framework for a
fundamental scheme in two space dimensions.

\section{Numerical   solid  boundary condition  } \label{cl-d2q9}

\begin{center}   \includegraphics[height=6.5cm]   {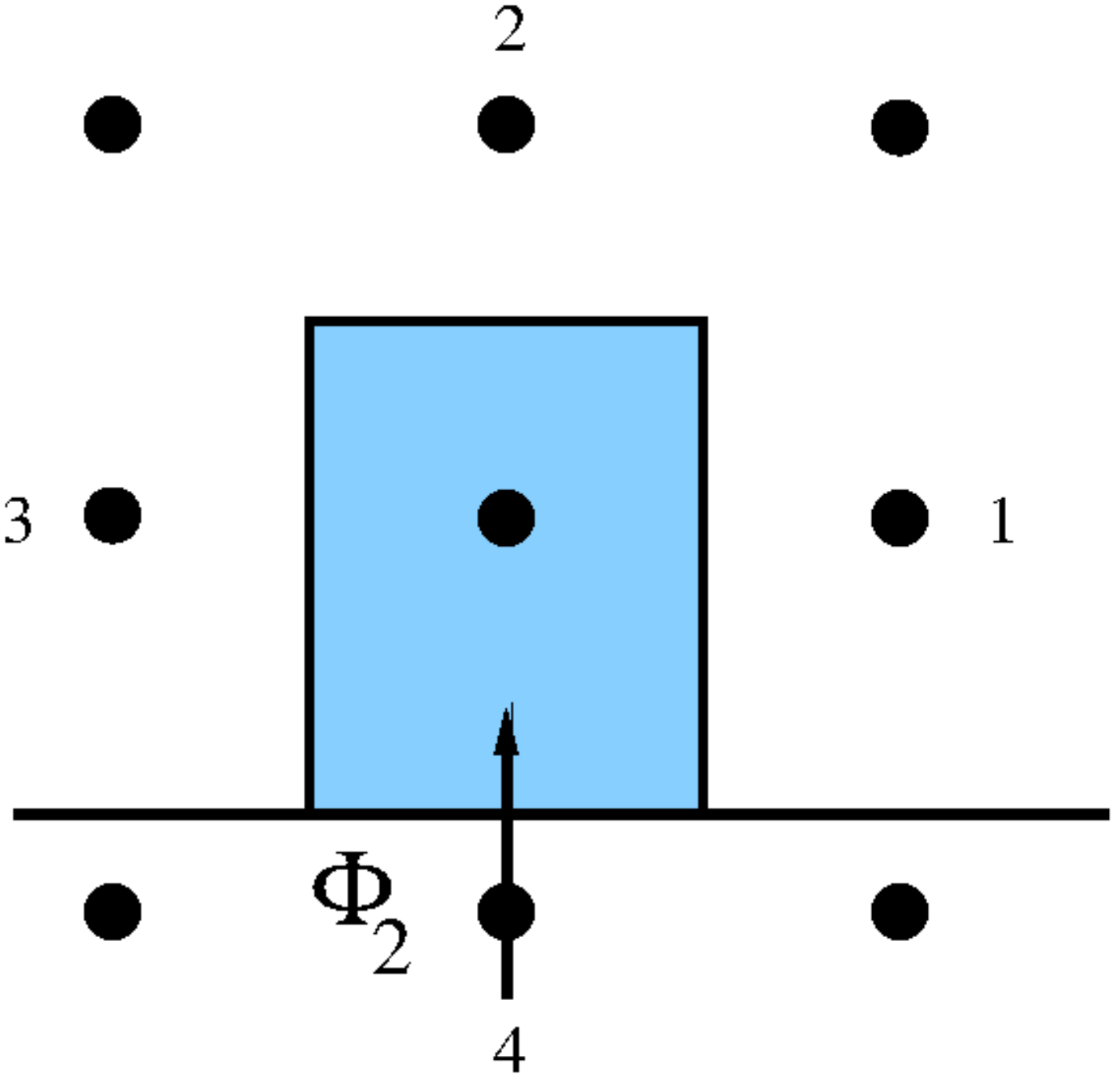}
\qquad \qquad   \includegraphics[height=6.5cm]   {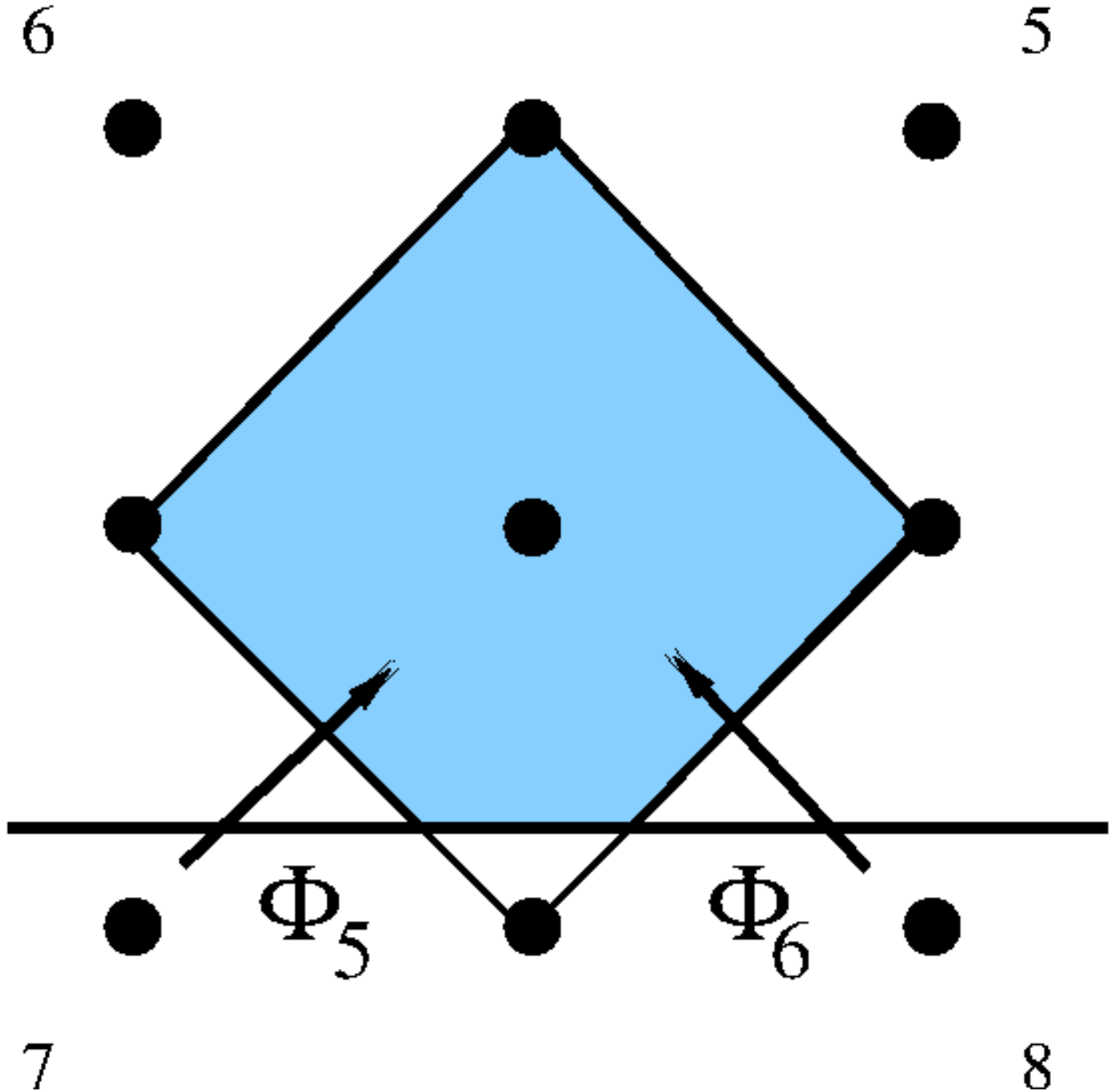} \end{center}
\begin{center} \it Figure 8. Control finite  volumes $ \, K _{\|} \,$  and   
$ \, K _{\times} \,$    near the boundary. 
 \end{center} \bigskip 

\nototo \monitem
We study in this section the example of an horizontal  
impenetrable  solid boundary for regular geometry
that is parallel to the axis of coordinates. We denote by $ \, x \equiv (x_1 ,\, x_2) \, $ 
a vertex located near the
boundary; the latter is in this contribution supposed to be parallel to the $x_1$ coordinate
axis

\moneq 
y_{\rm w} = x_2 - \xi \, \Delta x \, .
\label{frontiere-2D}  \end{equation}

\smallskip  \noindent 
A no-slip  boundary condition is supposed to be given for the fluid at the above 
boundary  (\ref{frontiere-2D}):

\moneq 
u ( \smb \,,\, y_{\rm w}) \equiv V  ( \smb)  \, , 
\label{vit-nulle}  \end{equation}

\smallskip  \noindent 
where $\, V  ( \smb) \, $ is some velocity field tangential to the boundary. 
Then, as illustrated in  Figure 8,  the number of  ``post-collision'' particles 
$ \, f_2^*(x-v_2 \Delta t) ,\, $ 
 $ \, f_5^*(x-v_5 \Delta t) \, $ and  $ \, f_6^*(x-v_6 \Delta t) \, $
coming from the neighbours $ \, x_4 ,$ $ \, x_7 $  and $ \, x_8 \, $ 
respectively of the node $x$ 
are not given by the general scheme (\ref{schema}).

 \monitem 
We denote by $ \, \vert K_{\|} \vert \,$ and  $ \, \vert K_{\times} \vert \,$ the measures 
of the finite volumes around the vertex $ \, x \, $ defined according to Figure 8. 
The boundary  $\, \partial  K_{\|} \,$ is composed by the  four edges $\, a_j \,$
for $j = 1$ to $ 4 $ and  $\, \partial  K_{\times } \,$ by the  {\bf five}  edges $\, a_j \,$
for $j = 5$ to $ 9  . \, $ 
Note in passing that the edge $\, a_9 \, $ is on the solid boundary. 
Instead of the relations  (\ref{vol-partiel}) and 
(\ref{mes-edges}), we have 

\moneq   
 \vert  K_{\|} \vert = \Big( {{1}\over{2}} + \xi \Big) \, \Delta x ^2 \,, \qquad  
\vert  K_{\times } \vert = \big( 1 + 2 \xi - \xi^2 \big) \, \Delta x ^2 \,. 
\label{vol-partiel-2D} \end{equation}

\moneq   
 \vert  a_1  \vert =  \vert  a_3  \vert = \Big( {{1}\over{2}} + \xi \Big) \, 
\Delta x  \,, \,\, 
 \vert  a_2  \vert =  \vert  a_4  \vert = \Delta x  \,, \,\, 
\label{par-edges-2D} \end{equation}

\moneq   
 \vert  a_5  \vert =  \vert  a_6 \vert =  \, \Delta x  \, \sqrt{2} \,, \,\, 
 \vert  a_7  \vert =  \vert  a_8  \vert = \xi \, \Delta x  \, \sqrt{2}  \,, \,\, 
 \vert  a_9  \vert = 2 \, (1 - \xi) \,  \Delta x \, . 
\label{croix-edges-2D} \end{equation}

\monitem 
{\bf Scheme 2. \quad  Flux  boundary condition  for the D2Q9 model.}

\noindent
We denote by  $ \Phi_2 \,, $ $ \, \Phi_5  \,$ and $ \, \Phi_6 $  the unknown 
incoming particle numbers. Recall that 

\moneq 
 \Phi_j   =  f_j^*(x-v_j \Delta t) \equiv  f_j^*(x_{\sigma (j)} ) \, , 
\qquad j = 2 \,,\, \, 5  \,,\, \, 6 \, .  
\label{def-phi}  \end{equation}

\smallskip  \noindent 
We use this  notation because the vertices  $ \, x_4 ,$ $ \, x_7 $  and 
$ \, x_8 \, $  are {\it not } defined as nodes of the computational domain. 
When we write the approximate  conservation of mass (\ref{evol-ro-paral}) 
 (\ref{evol-ro-croix}) in the volumes  
$\, \partial  K_{\|} \,$ and   $\, \partial  K_{\times } \,$ 
and the conservation of tangential momentum  (\ref{evol-q-paral})
 (\ref{evol-q-croix}) in the control volume 
 $\, \partial  K_{\times } ,  \,$ it is possible to define the 
three unknown particle distributions  $ \Phi_2 \,, \, \Phi_5  \,, \, \Phi_6 $  
according to

\moneq   
 \Phi_2  =  f_4^* - {{1 - 2 \, \xi}\over{1  + 2 \,\xi}} \Big( f_2^* - f_4^*(x_2) \Big) \,, 
\label{phi2-2D} \end{equation}

\smallskip  \noindent 
for the normal input particle number  across the boundary and to 

\moneq   
 \Phi_5  =  f_8^* + {{1 - \xi}\over{1 + \xi}} \, \Big( \! -f_5^* + f_8^*(x_6) \Big) - 
{{1}\over{\xi \, (1 + \xi)}} \, {{1}\over{{\cal R}_\Delta}}  \, {{\delta q_{\rm w}
 }\over{\lambda}} \, 
\label{phi5-2D} \end{equation}

\moneq   
 \Phi_6  =  f_7^* + {{1 - \xi}\over{1 + \xi}}  \, \Big( \! -f_6^* + f_7^*(x_5) \Big) + 
{{1}\over{\xi \, (1 + \xi)}} \, {{1}\over{{\cal R}_\Delta}}  \, {{\delta q_{\rm w}
 }\over{\lambda}} \,  
\label{phi6-2D} \end{equation}

\smallskip  \noindent  
for the transverse  input particle numbers, with 
   $\, {\cal R}_\Delta \,$ and $\, \delta q_{\rm w} \,$ 
defined according to 

\moneq   
\delta q_{\rm w} =  \sum_{j=0}^8 v_j^x \, f_j(x) - \rho \, V \,
\label{dq} \end{equation}

\moneq   
{\cal R}_\Delta \equiv  {{ \rho \, \lambda \, \Delta x }\over { \mu }} \,. 
\label{reynolds} \end{equation}

 \smallskip  \noindent
and  $ \,\lambda \,$ introduced in  (\ref{lambda}).

\monitem 
Note that this kind of truly two-dimensional treatment is unusual in the 
framework of lattice Boltzmann schemes, except for the pioneering work of \cite{mbg96}. 
We remark that the incoming particle distributions $\, \Phi_2 ,\, $  $\, \Phi_5 ,\, $ 
and  $\, \Phi_6 \, $  are expressed as {\bf linear}  functions of the other internal particle 
distributions  $\,  f_j^*(x_k) \,$ and of the boundary data. This is due to the fact that our
methodology is essentially based on the conservation laws 
of mass and momentum that are linear in terms of conserved 
variables and fluxes. All the nonlinearities are taken in consideration through 
the collision step $ \, f \longrightarrow f^* . \, $

\smallskip \noindent {\bf Construction of Scheme 2.}

\smallskip \noindent
We first explain the  notations used in relations  (\ref{phi5-2D}) and 
(\ref{phi6-2D}). First, according to the classical form of the Navier Stokes equations 
\cite{ll59} and to the hypothesis of an inpenetrable boundary,
the tangential flux $\, \tau \, $ across the edge $\, a_9 \,$ 
is defined in terms of the viscosity $\, \mu \,$ and the normal derivative 
$\, {{\partial u^x}\over{\partial n}} \, $ of the tangential velocity: 

\moneq   
 \tau = - \mu \,  {{\partial u^x}\over{\partial n}} \,. 
\label{def-tau} \end{equation}
 
\smallskip  \noindent 
For  the particular case we study  in this contribution,  the normal $ \, n \, $ is pointing
in the negative $y$ direction. We approximate  $\,- {{\partial u^x}\over{\partial n}} \, $ 
by a two-point finite difference scheme using the tangential momentum
$ \,  \sum_{j=0}^8 v_j^x \, f_j(x) \,$ and the (supposed to be constant) reference density
$ \, \rho \, $  at the vertex $x$. Then we have

 \smallskip  \noindent $  \displaystyle
 \tau =  \mu \,  {{ u^x(x) - V }\over{\xi \, \Delta x }} \, = \, 
{{\mu}\over{\rho}} \,   {{1}\over{\xi \, \Delta x }} \, 
\Big( \sum_{j=0}^8 v_j^x \, f_j(x) - \rho \, V \Big)  \, . \,$ 

 \smallskip  \noindent
It is then natural to consider the difference of tangential momentum 
$\, \delta q_{\rm w}  \,$    (defined in (\ref{dq}))  between the
computed value at the vertex $ \, x \, $ and the given value on the (wall) boundary 
 and the grid Reynolds number $ \, {\cal R}_\Delta \, $  (defined in (\ref{reynolds}))
associated with the mesh speed $\, \lambda \, $  and the space
increment $ \, \Delta x . \, $   With these notations, we have 

\moneq   
 \tau =  {{\lambda }\over{\xi \, {\cal R}_\Delta}} \, \delta q_{\rm w} \, . 
\label{tau} \end{equation}

\monitem
We write now the conservation  (\ref{evol-ro-paral})  of partial  mass $\,  \rho_{\|} .\, $ 
First due to the Boltzmann scheme

\moneq     
 \rho_{\|}(x,\, t + \Delta t)  -  \rho_{\|}^* (x,\,  t)  
  +   \sum_{j=1}^3   \big(    f^*_j (x)  -  f^*_{\sigma(j)}  ( x_j )  \big)  
 +  \big(    f^*_4  (x) - \Phi_2 \big)  =  0  \,.
\label{evol-ropar-1} \end{equation} 

 \smallskip  \noindent
Second due to  the conservation  (\ref{evol-ro-paral})  inside the volume  $\, K _{\|} :\, $ 

\moneq  
 {{1}\over{\Delta t}} \, \big(  \rho_{\|}(x,\, t + \Delta t)  -
 \rho_{\|}^* (x,\,  t)  \big)   +   \sum_{j=1}^3    
 {{\vert  a_j \vert }\over{\vert  K _{\|} \vert}} \, 
\lambda \,  \big(    f^*_j (x)  -  f^*_{\sigma(j)}  ( x_j )  \big)  =  0 \,, 
\label{evol-ropar-2} \end{equation} 

 \smallskip  \noindent
making use of  the fact that the mass flux across the boundary $\, a_4 \,$ is {\bf null}. 
We eliminate the quantity $\,  (  \rho_{\|}(x,\, t + \Delta t)  -
 \rho_{\|}^* (x,\,  t) ) \,$ between the relations (\ref{evol-ropar-1}) and 
 (\ref{evol-ropar-2}) with the help of  the geometrical lemmas (\ref{vol-partiel-2D}) and 
(\ref{par-edges-2D}). Then the relation (\ref{phi2-2D})   is straightforward to derive. 
We observe that in the ``regular'' case when $ \, \xi = {{1}\over{2}}, \,$ we recover the
  ``bounce-back'' boundary condition.

\monitem
In a similar way, we write  the conservation  (\ref{evol-ro-croix}) of partial 
 mass $\,  \rho_{\times} \, $  first due to the Boltzmann scheme

\moneq   \begin{cases}  \displaystyle   
\rho_{\times}(x,\, t + \Delta t)  -  \rho_{\times}^* (x,\,  t)  
+ \sum_{j=5}^6   \big(    f^*_j (x,\,t)  -  f^*_{\sigma(j)}  ( x_j ,\, t )  \big)  +  \\ 
 \displaystyle  \qquad   \qquad   \qquad 
  +  \big(    f^*_7  (x,\,t) - \Phi_5 \big)  +  \big(    f^*_8  (x,\,t) - \Phi_6 \big)  =  0
\end{cases} \label{evol-rocroix-1} \end{equation} 

\smallskip  \noindent
and second according to the mass conservation  (\ref{evol-ro-croix}) 
 inside the volume  $\, K_{\times} :\, $ 

\moneq 
  \begin{cases}  \displaystyle    \qquad  
 {{1}\over{\Delta t}}  \big(  \rho_{\times}(x,\, t + \Delta t)  -
 \rho_{\times}^* (x,\,  t)  \big) + 
\\   \displaystyle  \qquad   \qquad   \qquad  
   +    \sum_{j=5}^6   
   {{\vert  a_j \vert \,  \lambda \, \sqrt{2}  }\over{\vert  K _{\times}  \vert   }}  
\,  \Big(    f^*_j (x)  -  f^*_{\sigma(j)}  ( x_j ) \Big)  + 
\\   \displaystyle  \quad  
+    {{ \lambda \,  \sqrt{2}   }\over{\vert   K _{\times}  \vert   }} 
\Big(  \vert  a_7 \vert \,    \big(    f^*_7  (x,\,t) - \Phi_5 \big)  +  
\vert  a_8 \vert \,     \big(    f^*_8  (x,\,t) - \Phi_6 \big) \Big)   =  0 \,. 
\end{cases} \label{evol-rocroix-2} \end{equation} 

\smallskip  \noindent
Once again, the fact that there is no mass flux across the edge $\, a_9 \,$ 
expresses  the physical boundary condition.  Then by elimination of 
$ \,  \big(  \rho_{\times}(x,\, t + \Delta t)  -  \rho_{\times}^* (x,\,  t)  \big) \, $ 
between the relations 
 (\ref{evol-rocroix-1}) and  (\ref{evol-rocroix-2}),  we obtain:

 \smallskip  \noindent $  \displaystyle 
 \sum_{j=5}^6    \Big(  1 - {{ \vert a_j \vert\,  \lambda \, \sqrt{2} \, \Delta t}
\over { \vert K_{\times} \vert }}   \Big) 
\,  \big( f^*_j (x)  -  f^*_{\sigma(j)}  ( x_j ) \big) $ 

 \smallskip  \noindent $  \displaystyle \quad +  
 \Big(  1 - {{ \vert a_7 \vert\,  \lambda \, \sqrt{2} \, \Delta t}
\over { \vert K_{\times} \vert }}   \Big) 
\,  \big( f^*_7 (x)  -  \Phi_5  \big) \,+\,  
\Big(  1 - {{ \vert a_8 \vert\,  \lambda \, \sqrt{2} \, \Delta t}
\over { \vert K_{\times} \vert }}   \Big) 
\,  \big( f^*_8 (x)  -  \Phi_6  \big) \,=\, 0  $

\smallskip  \noindent
Due to (\ref{vol-partiel-2D}) and  (\ref{croix-edges-2D}), we have

\moneq   
  1 - {{ \vert a_5 \vert\,  \lambda \, \sqrt{2} \, \Delta t}
\over { \vert K_{\times} \vert }}   \,=\, 
  1 - {{ \vert a_6 \vert\,  \lambda \, \sqrt{2} \, \Delta t}
\over { \vert K_{\times} \vert }}  \,=\, - {{(1 - \xi)^2}\over{1 + 2\, \xi - \xi^2}} 
 \label{coef-56} \end{equation} 

\moneq   
  1 - {{ \vert a_7 \vert\,  \lambda \, \sqrt{2} \, \Delta t}
\over { \vert K_{\times} \vert }}   \,=\, 
  1 - {{ \vert a_8 \vert\,  \lambda \, \sqrt{2} \, \Delta t}
\over { \vert K_{\times} \vert }}  \,=\,  {{(1 - \xi^2)}\over{1 + 2\, \xi - \xi^2}} \, . 
 \label{coef-78} \end{equation}

\smallskip  \noindent
Then 

 \smallskip  \noindent $  \displaystyle \quad  
-(1-\xi) \,  \big(  f^*_5 +    f^*_6 - f^*_7(x_5) -  f^*_8(x_6) \big)
\,+\, (1+\xi)  \,   \big(  f^*_7 -  \Phi_5 +  f^*_8 -  \Phi_6  \big)  \,=\, 0 $

 \smallskip  \noindent
We deduce an expression for the sum $ \,  \Phi_5 +  \Phi_6 : $ 

\moneq     
 \Phi_5 +  \Phi_6 =  f^*_7 +    f^*_8 - {{1-\xi}\over{1 + \xi }} \, \big( 
  f^*_5 +    f^*_6 - f^*_7(x_5) -  f^*_8(x_6) \big) \, . 
 \label{somme} \end{equation} 

\monitem
We now carefully express the conservation of tangential momentum. As in the previous
cases, we first have the expression directly  derived  from the scheme 
 (\ref{evol-q-croix}) 

\moneq   \begin{cases}  \displaystyle  
   q_{\times}^x (x,\, t + \Delta t)  -  q_{\times}^{*,\,x} (x,\,  t)
  +   \sum_{j=5}^6  v_{j}^x \, \big(    f^*_j (x)  +  f^*_{\sigma(j)}  ( x_j )
  \big) \,  +  
 \\   \displaystyle  \qquad   \qquad   \qquad 
  +  \, v_{7}^x \,   \big(    f^*_7  (x,\,t) + \Phi_5 \big)  +  
v_{8}^x \,  \big(    f^*_8  (x,\,t) + \Phi_6 \big)  =  0
\end{cases} \label{evol-q-1} \end{equation} 

\smallskip  \noindent
and second we have the (approximate !) conservation  (\ref{evol-q-croix})  
of tangential momentum inside the volume 
 $ \,  K _{\times} \, $

\moneq  
 {{1}\over{\Delta t}}   \big( q_{\times}^x(x,\, t + \Delta t)  -  q_{\times}^{*,\,x} (x,\,  t)
 \big) +    {{1}\over{\vert  K _{\times}  \vert   }}  \sum_{j=5}^9  \vert  a_j    \vert  
\,  \zeta_j^x = 0 \, . 
\label{evol-q-2} \end{equation} 

\smallskip \noindent
Due to   (\ref{zeta-croix}) and the  expressions of tangential speeds for the
D2Q9 model, {\it id est}

\moneq  
v_5^x =  \lambda \,, \quad v_6^x = - \lambda \,, \quad 
v_7^x =  -\lambda \,, \quad v_8^x =  \lambda \,, 
\label{vitesses} \end{equation} 

\smallskip \noindent
we have 
 
\moneq   \begin{cases}  \displaystyle    \,\,
  \zeta_5^x =  \lambda \,  v_5^x \,  \sqrt{2}\, \Big( f_5^* + f_7^*(x_5) \Big) \,, \quad 
  \zeta_6^x =  \lambda  \,  v_6^x  \, \sqrt{2}\, \Big( f_6^* + f_8^*(x_6) \Big) \,, 
 \\   \displaystyle    \,\,
  \zeta_7^x =  \lambda  \,  v_7^x  \, \sqrt{2}\, \Big( f_7^* + \Phi_5 \Big) \,, \qquad 
  \zeta_8^x =  \lambda  \,  v_8^x  \, \sqrt{2}\, \Big( f_8^* + \Phi_6 \Big) \, . 
\end{cases}  \label{zeta-x} \end{equation} 

\smallskip \noindent
The last term $ \, \zeta_9^x \,$  corresponds to the stress tensor along the (little)
cut edge  $ \,  a_9 \,$ which is the fifth edge of control volume  
 $ \,  K _{\times} \, $  as presented in  (\ref{croix-edges-2D}) (see also Figure 8). Then 
we have simply  

\moneq  
 \zeta_9^x = \tau \,   . 
\label{zeta-9} \end{equation}

\smallskip \noindent
We eliminate  $ \, ( q_{\times}^x(x,\, t + \Delta t)  -  q_{\times}^{*,\,x} (x,\,  t) ) 
\, $  between the relations (\ref{evol-q-1}) and  (\ref{evol-q-2}):

 \smallskip  \noindent $  \displaystyle 
 \sum_{j=5}^6    \Big(  1 - {{ \vert a_j \vert\,  \lambda \, \sqrt{2} \, \Delta t}
\over { \vert K_{\times} \vert }}   \Big) \, v_j^x \, 
\,  \big( f^*_j (x)  +  f^*_{\sigma(j)}  ( x_j ) \big) \, + $ 

 \smallskip  \noindent $  \displaystyle \qquad +  
 \Big(  1 - {{ \vert a_7 \vert\,  \lambda \, \sqrt{2} \, \Delta t}
\over { \vert K_{\times} \vert }}   \Big)  \, v_7^x 
\,  \big( f^*_7 (x)  +  \Phi_5  \big) \,+\, $

 \smallskip  \noindent $  \displaystyle \qquad   \qquad + 
 \Big(  1 - {{ \vert a_8 \vert\,  \lambda \, \sqrt{2} \, \Delta t}
\over { \vert K_{\times} \vert }}   \Big)  \, v_8^x 
\,  \big( f^*_8 (x)  +  \Phi_6  \big) \,=\, 
 {{ \vert a_9 \vert\,  \Delta t} \over { \vert K_{\times} \vert }} \, \tau    $

 \smallskip  \noindent
Due to   (\ref{vol-partiel-2D}), (\ref{coef-56}) and  (\ref{coef-78}), we have  

 \smallskip  \noindent $  \displaystyle 
-(1-\xi)^2 \,\Big( \lambda \, \big(  f^*_5 + f^*_7(x_5) \big) - 
 \lambda \, \big(  f^*_6 + f^*_8(x_6) \big)  \Big)  + $ 

 \smallskip  \noindent $  \displaystyle \qquad +  
(1-\xi^2) \, \Big( - \lambda \, \big(  f^*_7 +    \Phi_5  \big) + 
 \lambda \, \big(  f^*_8 +    \Phi_6  \big)   \Big)  \,=\, 2 \, (1-\xi) \Delta x 
\, {{\Delta t }\over{\Delta x^2}} \,  {{\lambda }\over{\xi \, {\cal R}_\Delta}} \, 
\delta q_{\rm w} \, . $

\smallskip \noindent
We divide the previous expression by $\, \lambda \, (1-\xi) \,$ and we deduce 

 \smallskip  \noindent $  \displaystyle 
(1+\xi) \,  \Big(   \Phi_5 -  \Phi_6  \Big) =  -(1-\xi) \, 
 \Big(  f^*_5 - f^*_6 + f^*_7(x_5) -  f^*_8(x_6)  \Big)   \,$ 

 \smallskip  \noindent $  \displaystyle  \qquad  \qquad  \qquad  \qquad  \qquad   \qquad   - 
(1+\xi) \,   \Big(  f^*_7 -  f^*_8  \Big) - 2 \, 
{{\delta q_{\rm w} }\over{ \lambda \, \xi \, {\cal R}_\Delta}} \,. $  

\smallskip \noindent
In other terms:

\moneq   \begin{cases}  \displaystyle 
  \Phi_5 -  \Phi_6 =  -f^*_7 +    f^*_8 - {{1-\xi}\over{1+\xi}} \, 
         \Big(  f^*_5 - f^*_6 + f^*_7(x_5) -  f^*_8(x_6) \Big) 
\\   \displaystyle  \qquad  \qquad   \qquad  \qquad  \qquad  \qquad 
-    {{2}\over{1+\xi}}  \,  {{1}\over{\xi}} \,  
{{ \delta q_{\rm w} }\over { {\cal R}_\Delta \, \lambda }}  \,. 
\end{cases}   \label{diff} \end{equation} 

\smallskip \noindent
The relations (\ref{phi5-2D}) and  (\ref{phi6-2D})  are 
obtained from    (\ref{somme}) and   (\ref{diff})
by the resolution of a two by two linear system.   $\hfill \square$

\bigskip 
\bigskip

\monitem 
{\bf  Couette  test case.}

 \bigskip  
\begin{center}   \includegraphics[height=6.5cm]    {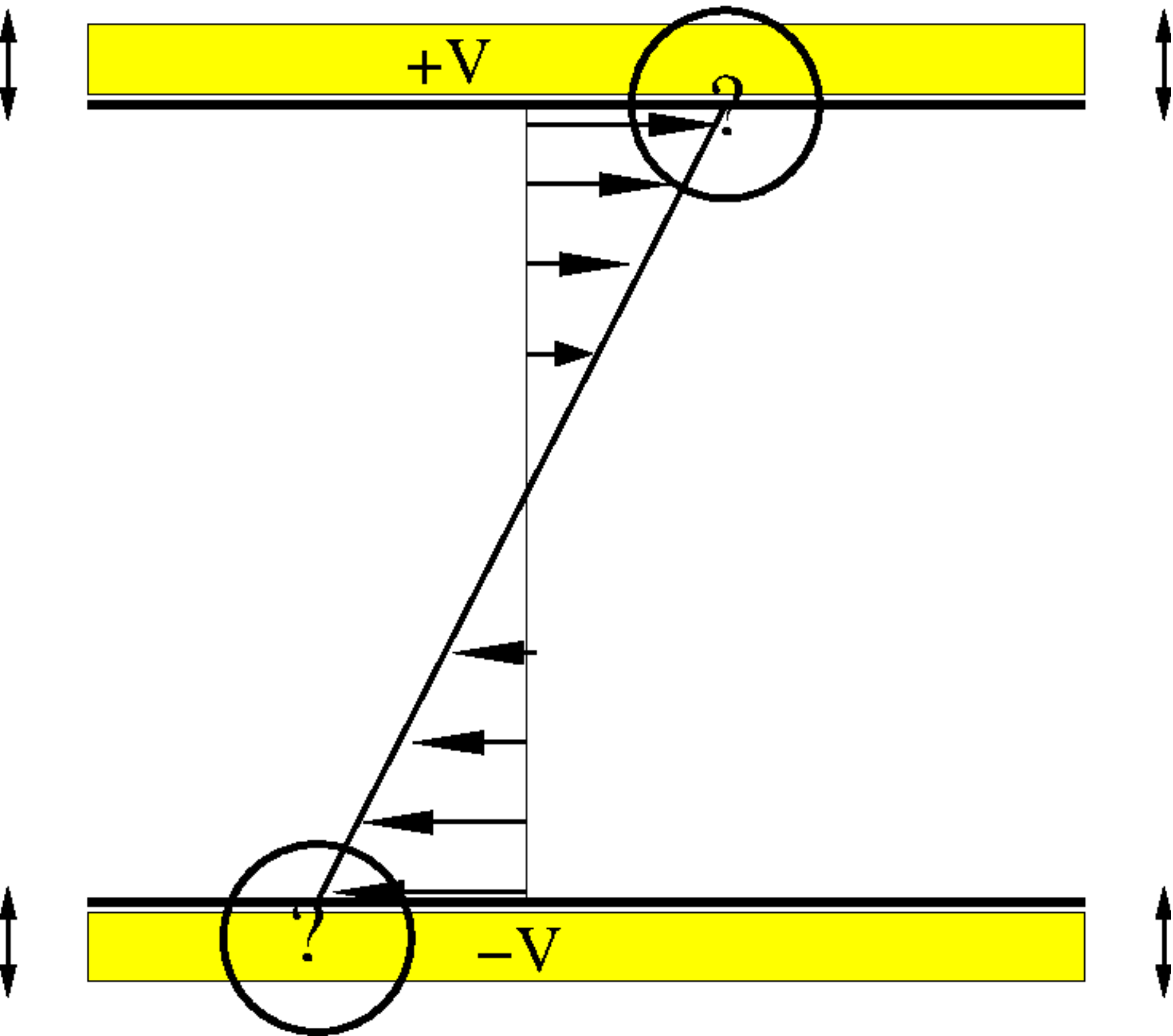}    \end{center}
\begin{center} \it Figure 9. Typical Couette flow.
 \end{center} \bigskip 

 \noindent
This classical flow is described in Figure 9. The boundary conditions are simply $+V$ on
top and $-V$ at the bottom of a channel. 
We have used several schemes proposed by 
D.~ D'Humi\`eres \cite{dh01}, Bouzidi {\it et al}  \cite{bfl01}, Ginzburg and  D'Humi\`eres 
 \cite{gh03} for a mesh composed by only 11 mesh points in the direction transverse to the
 flow.  We vary the location of the
 physical boundary in some proportion $\, \xi \,$ relatively to the mesh step $\, \Delta x
 .\,$  We compute the stationary discrete solution of our lattice Boltzmann scheme. Then
 with a linear regression fit we measure the location of the point associated with an
 extrapolated velocity exactly equal to  $+V$ or $-V$. Up to seven decimals, all the
 boundary schemes give the desired result of $ +  \xi \, \Delta x $ or  $ \, -  \xi \,
 \Delta x $.

 \bigskip 
\monitem 
{\bf  Poiseuille test case.}

 \noindent 
To test the proposed formulae for boundary conditions, we have also considered the 
simple Poiseuille flow with two boundaries parallel to the $Ox$ axis located
respectively at $y_1=(1-\xi)\Delta x$ and $y_2=(N_y + \xi) \Delta x$ as described in
Figure 10. The
flow is driven by applying a uniform internal force $\delta f$, such that
the velocity distribution should be of parabolic form, with null values for
$y_1$ and $y_2$ and a maximum value 
$v_m=\delta f (N_y-1+2 \xi)^2 (\Delta x)^2 \rho /(8 \mu)$.

\bigskip  
\begin{center}   \includegraphics[height=6.5cm]    {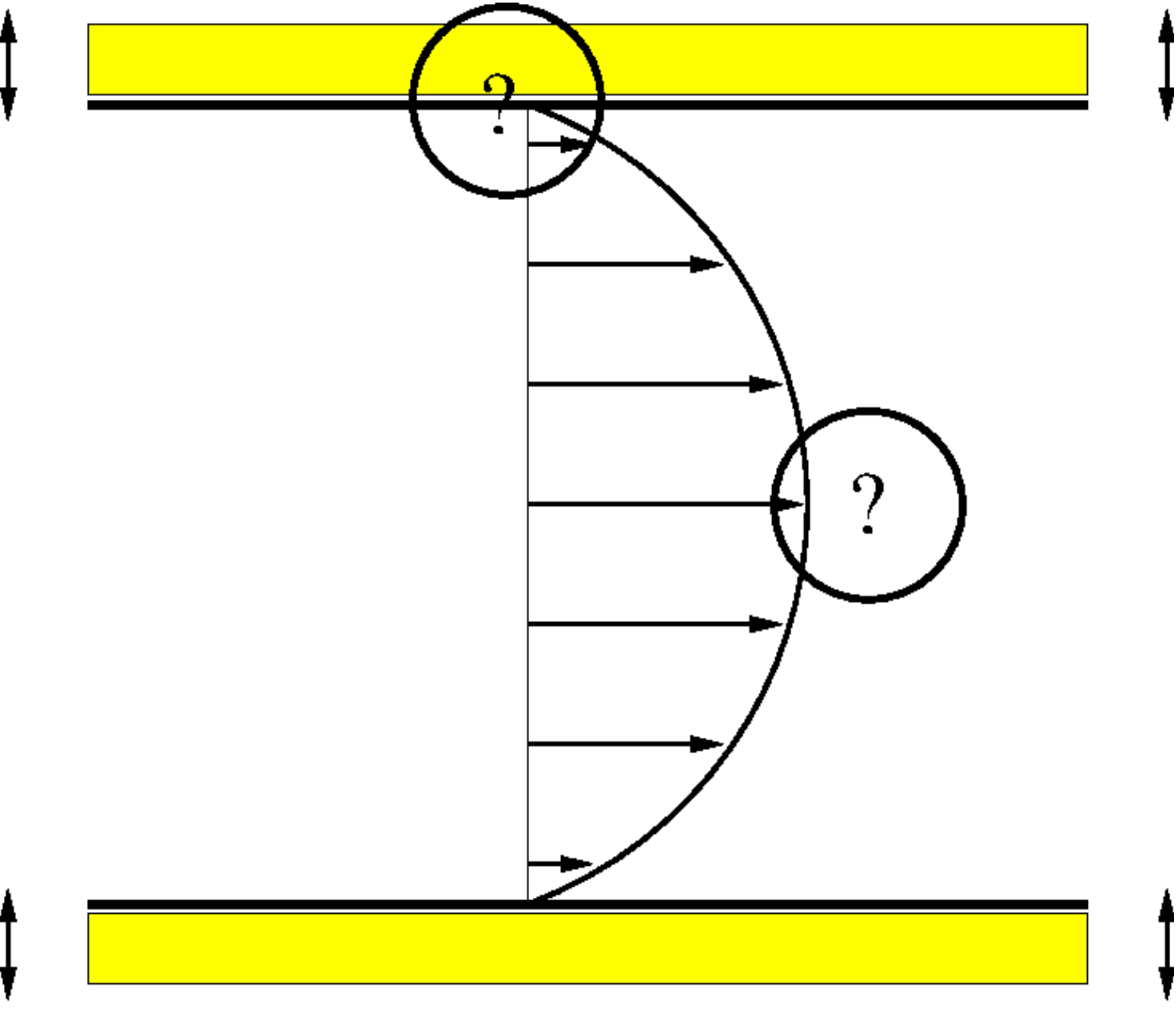}    \end{center}
\begin{center} \it Figure 10. Typical Poiseuille profile 

and quantities chosen to compare   model and theory. 
 \end{center} \bigskip 

\bigskip 

\noindent 

\setbox20=\hbox{ $\,\,$ }
\setbox30=\hbox{ 11 points }
\setbox50=\hbox{ 21 points }
\setbox60=\hbox{ 31 points }
\setbox21=\hbox{ DH }
\setbox31=\hbox{ $ 2.75397 \times 10^{-2} $ }
\setbox51=\hbox{ $ 8.1936  \times 10^{-3} $ }
\setbox61=\hbox{ $ 3.8728  \times 10^{-3} $ }
\setbox22=\hbox{ BFL1 }
\setbox32=\hbox{ $ 2.5873  \times 10^{-2} $ }
\setbox52=\hbox{ $ 7.6978  \times 10^{-3} $ }
\setbox62=\hbox{ $ 3.6384  \times 10^{-3} $ }
\setbox23=\hbox{ BFL2 }
\setbox33=\hbox{ $ 7.9811  \times 10^{-3} $ }
\setbox53=\hbox{ $ 2.1898  \times 10^{-3} $ }
\setbox63=\hbox{ $ 1.0049  \times 10^{-3} $ }
\setbox24=\hbox{ IGDH }
\setbox34=\hbox{ 0  }
\setbox54=\hbox{ 0  }
 \setbox64=\hbox{ 0  }
\setbox25=\hbox{ DL }
\setbox35=\hbox{ 0 }
\setbox55=\hbox{ 0 }
\setbox65=\hbox{ 0 }
\setbox26=\hbox{ D1 }
\setbox36=\hbox{ 1.0555950 }
\setbox56=\hbox{ 1.0168517 }
\setbox66=\hbox{ 1.0088778 }
\setbox27=\hbox{ DL0 }
\setbox37=\hbox{ 1  }
\setbox57=\hbox{ 1  }
\setbox67=\hbox{ 1  }
\setbox28=\hbox{ DL1 }
\setbox38=\hbox{ 1.0555556 }
\setbox58=\hbox{ 1.0165289 }
\setbox68=\hbox{ 1.0078125 }
\setbox44=\vbox{\offinterlineskip  \halign {
&\tvg#& # &\tvg#&   #  &\tvg#&  #  &\tvg#&  # &\tvd#\cr 
\na&  \box20 && \box30 &&  \box50  && \box60  \hcr 
\na&  \box21 && \box31 &&  \box51  && \box61  \hcr 
\na&  \box22 && \box32 &&  \box52  && \box62  \hcr 
\na&  \box23 && \box33 &&  \box53  && \box63  \hcr 
\na&  \box24 && \box34 &&  \box54  && \box64  \hcr 
\na&  \box25 && \box35 &&  \box55  && \box65  \hcr 
 \na}   }  \centerline{\box44  }

\begin{center} \it Table 3.    Largest discrepancy of the 
variation of the maximal velocity for a Poiseuille
  profile for several boundary schemes and meshes.  \end{center}

\begin{center}   \includegraphics [height=8.5cm]  
{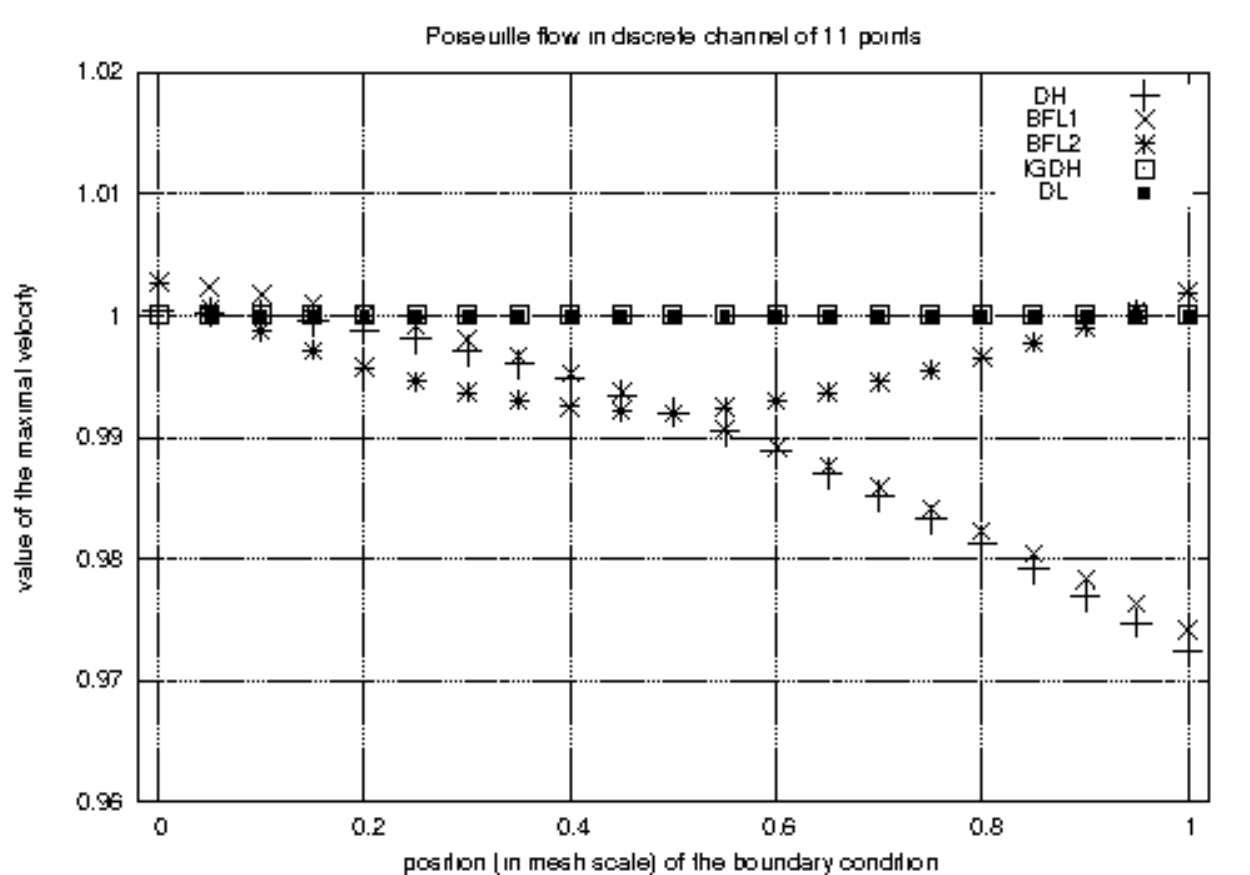} \end{center}
\begin{center} \it Figure 11    Variation of the maximal velocity for a Poiseuille
  profile.     \end{center}
\bigskip 

\bigskip 

 \monitem 
When steady state is reached in the simulation, the velocity profile is fit
to a parabolic distribution yielding  the
maximum velocity and the locations of 0-velocity from which an effective
channel width can be deduced.  
These values are compared to the theoretical values indicated
above. Note that when a driving force is applied, we follow \cite{gh03} and
perform the parabolic fitting with the quantity 
$(\sum v_j f_j )/\rho + 1/2 \delta f$. 
Data presented here correspond to the particular case $N_y=11$.
Figure 11 and Table 3 show results for the comparison 
of the measured maximum velocity normalized
by its theoretical value for various boundary schemes.
Similarly Figure 12 and Table 4 show the difference between the location of the lower point 
of 0-velocity and its imposed value {\it vs} $\xi$.
Obviously the simple bounce-back scheme which gives a constant location leads to an error
linear in $\xi$. 
 For comparison we show the results for a simple boundary condition and
indicate that an elaborate scheme, like that of Ginzburg and D'Humi\`eres 
gives the theoretical velocity profile to machine precision. 
We have also tested whether the proposed scheme statisfies Galilean invariance.
This is very well satisfied provided the expression for the equilibrium value
of the energy-squared moment includes a non-linear term $\, -6 \rho \, (j_x^2+j_y^2)$
that differ by a factor of $2$ from the term 
 provided by the simple BGK   equilibrium values \cite{qhl}.



 \bigskip  \bigskip 

\setbox20=\hbox{ $\,\,$ }
\setbox30=\hbox{ 11 points }
\setbox50=\hbox{ 21 points }
\setbox60=\hbox{ 31 points }
\setbox21=\hbox{ DH }
\setbox31=\hbox{ $ 8.31958  \times 10^{-2} $ }
\setbox51=\hbox{ $ 4.51576  \times 10^{-2} $ }
\setbox61=\hbox{ $ 3.10122  \times 10^{-2} $ }
\setbox22=\hbox{ BFL1 }
\setbox32=\hbox{ $ 7.81277  \times 10^{-2} $  }
\setbox52=\hbox{ $ 4.24195  \times 10^{-2} $ }
\setbox62=\hbox{ $ 2.91337  \times 10^{-2} $ }
\setbox23=\hbox{ BFL2 }
\setbox33=\hbox{ $ 2.19920  \times 10^{-2} $  }
\setbox53=\hbox{ $ 1.15029  \times 10^{-2} $ }
\setbox63=\hbox{ $ 7.79  \times 10^{-3} $  }
\setbox24=\hbox{ IGDH }
\setbox34=\hbox{ 0  }
\setbox54=\hbox{ 0  }
\setbox64=\hbox{ 0  }
\setbox25=\hbox{ DL }
\setbox35=\hbox{ 0 }
\setbox55=\hbox{ 0 }
\setbox65=\hbox{ 0 }
\setbox26=\hbox{ D1 }
\setbox36=\hbox{ 0.1642783 }
\setbox56=\hbox{ 0.0896240 }
\setbox66=\hbox{ 0.0590896 }
\setbox27=\hbox{ DL0 }
\setbox37=\hbox{ 0  }
\setbox57=\hbox{ 0  }
\setbox67=\hbox{ 0  }
\setbox28=\hbox{ DL1 }
\setbox38=\hbox{ 0.1644140 }
\setbox58=\hbox{ 0.0905365 }
\setbox68=\hbox{ 0.0623784 }
\setbox44=\vbox{\offinterlineskip  \halign {
&\tvg#& # &\tvg#&   #  &\tvg#&  #  &\tvg#&  # &\tvd#\cr 
\na&  \box20 && \box30 &&  \box50  && \box60  \hcr 
\na&  \box21 && \box31 &&  \box51  && \box61  \hcr 
\na&  \box22 && \box32 &&  \box52  && \box62  \hcr 
\na&  \box23 && \box33 &&  \box53  && \box63  \hcr 
\na&  \box24 && \box34 &&  \box54  && \box64  \hcr 
\na&  \box25 && \box35 &&  \box55  && \box65  \hcr 
 \na}   }  \centerline{\box44  }

\begin{center} \it Table 4.    Largest discrepancy of the 
variation of the point of zero velocity  for a Poiseuille
  profile for several boundary schemes and meshes.  \end{center}

\bigskip  
\begin{center}   \includegraphics   [height=8.5cm]  
{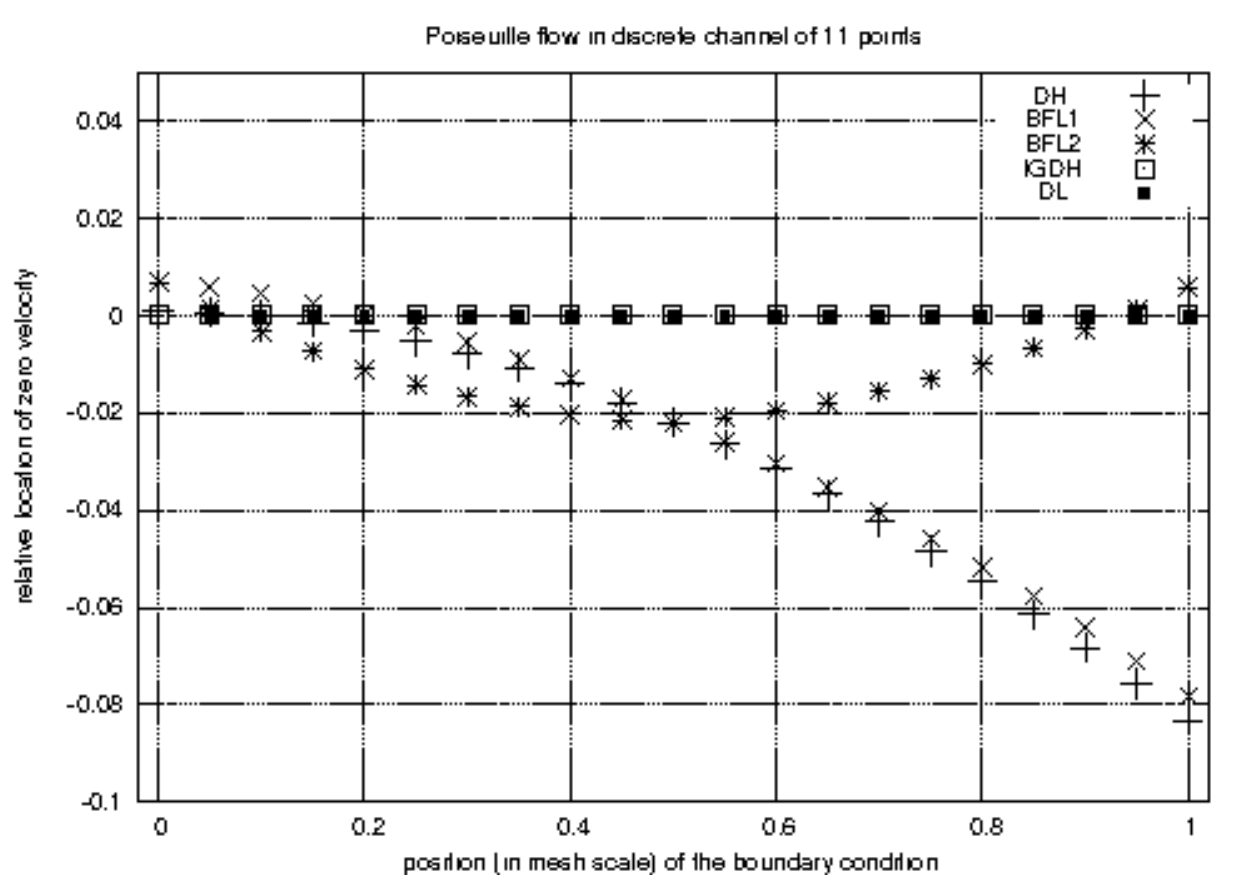} \end{center}
\begin{center} \it Figure 12   Location of the point of zero velocity for a Poiseuille
  flow.      \end{center}
\bigskip 

 \bigskip \vfill \eject
\monitem 
{\bf  Stokes eigenmode in a square domain }

\bigskip  
\begin{center} $\!\!\!\!\!\!\!\!\!\!\!\!\!\!\!\!\!\!\!\!\!\!\!\!\!\!\!\!$  
 \includegraphics [width=10.6cm, height=8.5cm]  
{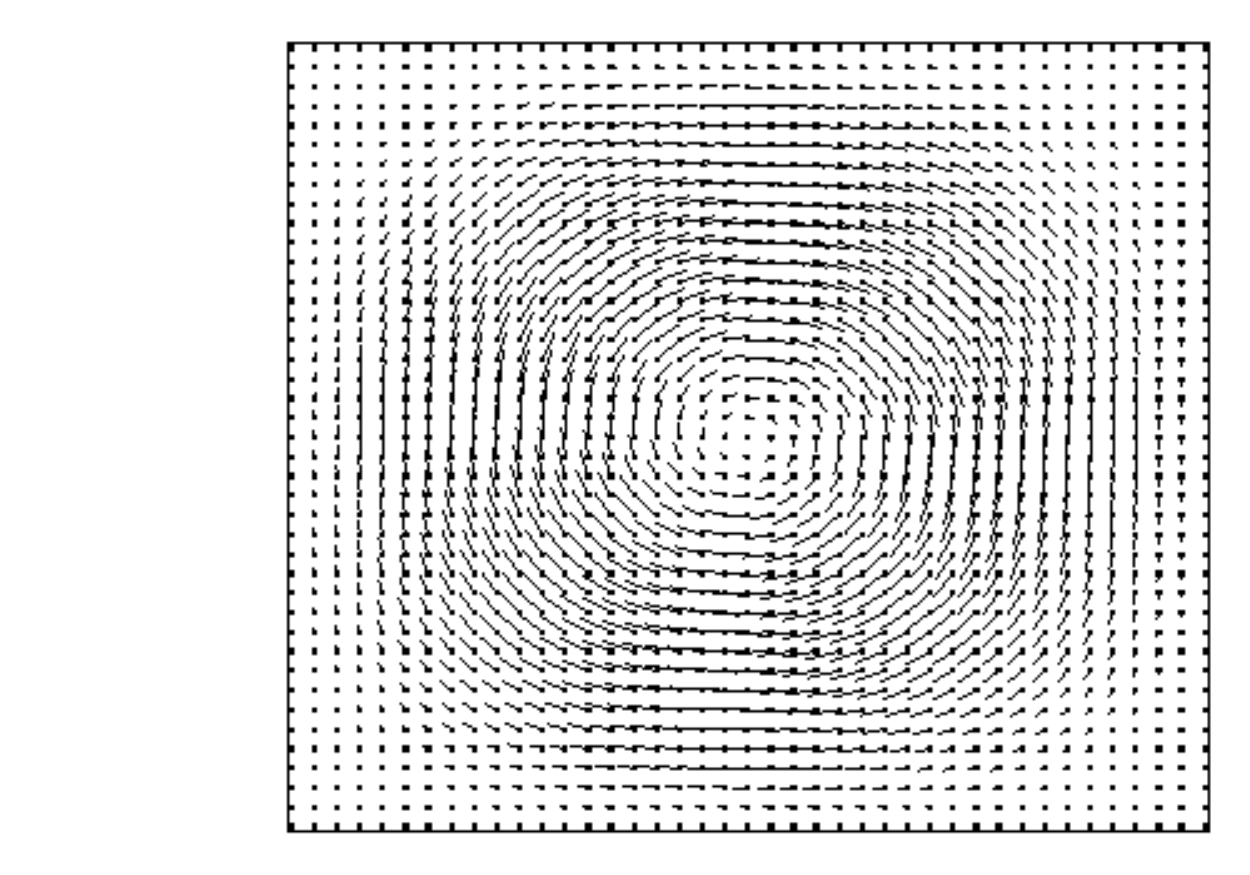}   \end{center}

\begin{center} \it Figure 13. First eigenfunction of the Stokes problem in
  a square. 
 
 \end{center} \bigskip 

 \noindent 
To perform a more significant test of the proposed boundary conditions, we
considered a simple but very well documented case, that of Stokes modes in a
square cavity  with homogeneous Dirichlet boundary conditions for velocity \cite{ll04}. 
Inside the cavity, the fluid follows Stokes equations, for this
we use D2Q9 with no non linear term for the equilibrium values of the 
non-conserved momenta 
 and set the relaxation rates such that there is no
 fourth order term in the equivalent equations. 
We use various boundary 
conditions to obtain zero velocity for the horizontal and vertical boundaries
of the square (for $x=1-\xi$ and $x=N+\xi$, and $y=1-\xi$ and $y=N+\xi$,
$0 < \xi <  1$),
so that the size of the square is $N-1+2 \xi$, with $N^2$ lattice nodes. The
values of the Stokes eigenmodes should scale as

\moneq  
\Gamma=\frac{\gamma(j) \, \nu}{(N-1+2 \xi)^2}
\end{equation}

 \smallskip  \noindent
where $\nu$ is the shear viscosity and
$\gamma(j)$ depends on the structure of the corresponding eigenmode
and is given for small values of $j$ by Labrosse {\it et al} \cite{ll04}.
We present in Figure 13 the vector field corresponding to  the first eigenvalue of the
Stokes problem. 
Using the Arnoldi procedure \cite{ar51}, we determine $\Gamma_{LB}$ for several values
of $N$ and plot in Figure~14  the relative error 
$ \, {{\Gamma_{LB}}\over{\Gamma}}-1 \, $ {\it vs} 
 $\, N^2\, $ for a few ways to
implement the boundary conditions. The reader can appreciate the quality
of the proposed boundary conditions. The data given in the Figure~14 correspond
to the lowest eigenmode, but similar behaviour is observed for higher order
modes (up to $j=30$).

\bigskip   \begin{center}   \includegraphics [height=8.5cm]  
{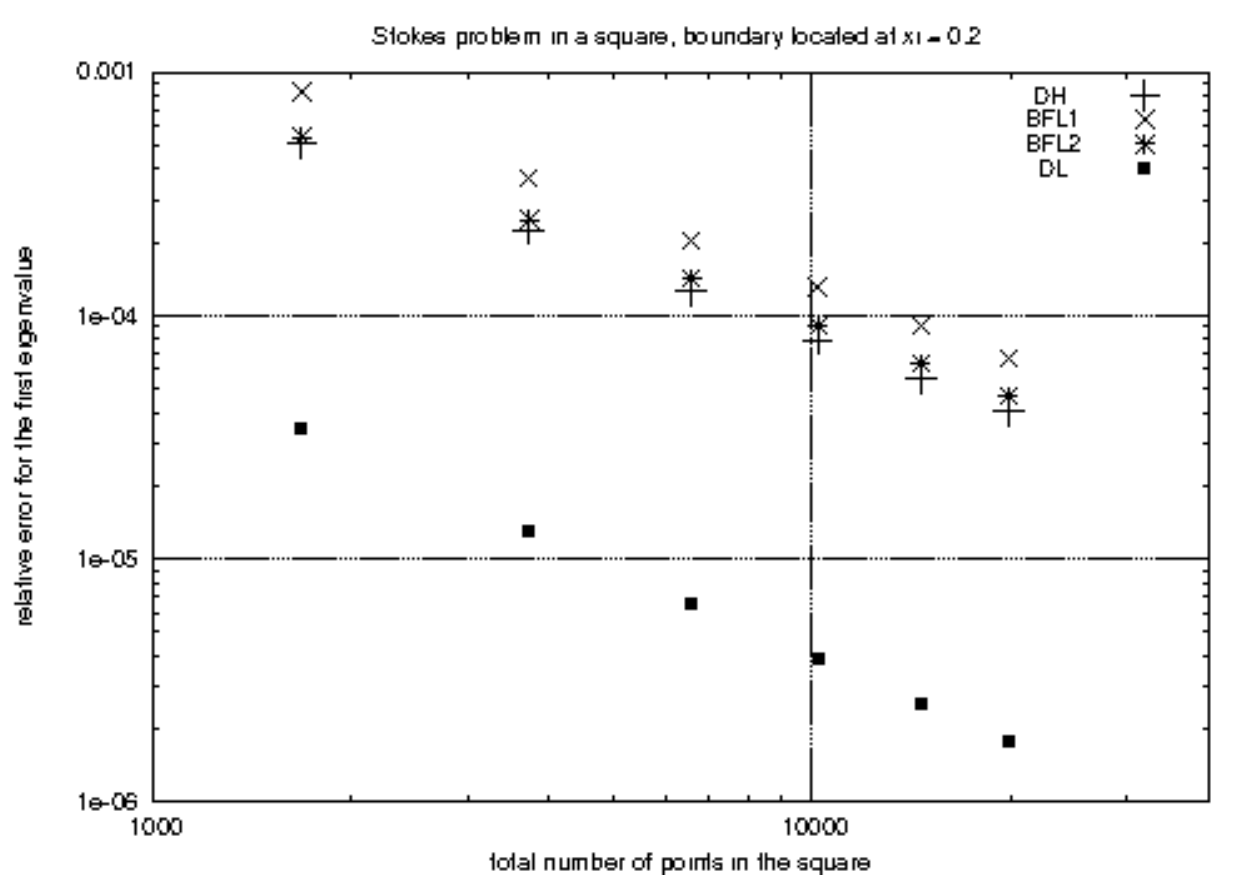} \end{center}

\begin{center} \it Figure 14. Discrepancies for  the first eigenvalue of the Stokes
  problem with D2Q9 lattice Boltzmann  scheme and various boundary conditions.
 \end{center}     \bigskip 

\section{Conclusion} \label{conclu}

\nototo \monitem
We have proposed a link between the lattice  Boltzmann scheme and the finite volume
method. In particular we proposed general relations that define mass and momentum fluxes
between two grid points of the lattice Boltzmann scheme. For the D2Q9 model, 
we have encountered geometrical difficulties and we have proposed the introduction of two
families of control volumes in order to define general mass and momentum fluxes. 
This approach naturally induces a flux methodology for the treatment of boundary
conditions when the boundary flux has naturally a physical meaning.
 Satisfactory  tests for  acoustic monodimensional wave, 
 solid two-dimensional boundary for a Couette  and Poiseuille flows, 
eigenmodes for the Stokes problem in a square   have been proposed. 
Our  boundary scheme appears to be very precise and can be conpared 
favorably with other   high accurate boundary schemes. 
The next step is to adapt the previous ideas for  a geometrically general stable 
algorithm for two-dimensional  boundary  conditions. An other extension concerns a more
precise treatment of the internal mass and momentum exchanges between the different
control volumes. A link with the so-called ``reservoir method'' \cite{avcl02} should be
explored.

\section{Acknowledgments} \label{merci}

\nototo \monitem
We thank F. Alouges  \cite{al05} who suggested to one of us the existence
 of a possible natural link between the
lattice Boltzmann scheme and the finite volume method. We thank also the referees who
suggested several improvements from the original version of this contribution.

   \end{document}